\documentclass{PoS}

\usepackage{graphicx}
\usepackage{amsmath}
\usepackage{amsfonts}
\usepackage{subfigure}
\usepackage{wrapfig}
\usepackage{epsfig}

\usepackage{graphicx}

\usepackage{afterpage,float}

\usepackage{amssymb}

\newcommand\fverb{\setbox\fverbbox=\hbox\bgroup\verb}
\newcommand\fverbdo{\egroup\medskip\noindent%
			\fbox{\unhbox\fverbbox}\ }
\newcommand\fverbit{\egroup\item[\fbox{\unhbox\fverbbox}]}
\newbox\fverbbox

\newcommand{\be}{\begin{equation}}
\newcommand{\ee}{\end{equation}}
\newcommand{\bea}{\begin{eqnarray}}
\newcommand{\eea}{\end{eqnarray}}

\def\half{{\textstyle{1\over2}}}

\def\CL{{\cal L}}

\def\CV{{\cal V}}

\def\Tr{{\sf Tr}}

\def\bar{\overline}

\def\tilde{\widetilde}

\def\half{{\scriptstyle \raise.15ex\hbox{${1\over2}$}}}

\newcommand{\beq}{\begin{equation}}
\newcommand{\eeq}{\end{equation}}

\newcommand{\real}{\relax{\rm I\kern-.18em R}}

\newcommand{\tr}{\mbox{\,tr\,}}

\newcommand{\eff}{{\rm eff}}

\newcommand{\eps}{\epsilon}

\def\vek#1{{\bf #1}}

\title{Chiral symmetry breaking in nearly conformal gauge theories \thanks{Based on talks at the 
conference by J.~Kuti, D.~Nogradi, and C.R.~Schroeder. }}

\ShortTitle{Chiral symmetry breaking in nearly conformal gauge theories}

\author{Zolt\'an Fodor\\
        Department of Physics, University of Wuppertal\\
        Gau$\beta$strasse 20, D-42119, Germany\\
        Email: \email{fodor@bodri.elte.hu}}

\author{Kieran Holland\\
        Department of Physics, University of the Pacific\\
        3601 Pacific Ave, Stockton CA 95211, USA\\
        Email: \email{kholland@pacific.edu}}

\author{Julius Kuti \\
        Department of Physics 0319, University of California, San Diego\\
        9500 Gilman Drive, La Jolla, CA 92093, USA\\
        E-mail: \email{jkuti@ucsd.edu}}

\author{D\'aniel N\'ogr\'adi\\
        Department of Physics 0319, University of California, San Diego\\
        9500 Gilman Drive, La Jolla, CA 92093, USA\\
        Email: \email{nogradi@bodri.elte.hu}}

\author{Chris Schroeder\\
        Department of Physics 0319, University of California, San Diego\\
        9500 Gilman Drive, La Jolla, CA 92093, USA\\
        E-mail: \email{crs@physics.ucsd.edu}}

\abstract{
We present new results on chiral symmetry breaking in nearly conformal gauge
theories with fermions in the fundamental representation of the SU(3) color
gauge group. The number of fermion flavors is varied in an extended range below
the conformal window with chiral symmetry breaking ($\chi{\rm SB}$) for all
flavors between $N_f=4$ and $N_f=12$.  To identify $\chi{\rm SB}$ we apply
several methods which include, within the framework of chiral perturbation
theory, the analysis of the Goldstone  spectrum in the p-regime and the
spectrum of the fermion Dirac operator with eigenvalue distributions of random
matrix theory in the $\eps$-regime. Chiral condensate enhancement is observed
with increasing $N_f$ when the electroweak symmetry breaking scale $F$ is held
fixed in technicolor language. Important finite-volume consistency checks from
the theoretical understanding of the $SU(N_f)$ rotator spectrum of the
$\delta$-regime are discussed. We also consider these gauge theories at
$N_f=16$ inside the conformal window. The importance of understanding
finite volume, zero momentum gauge field dynamics inside the conformal window is
pointed out.  Staggered lattice fermions with supressed taste breaking are used
throughout the simulations.
\vskip 0.05 in
}

\FullConference{The XXVII International Symposium on Lattice Field Theory - LAT2009\\
		 July 26-31 2009\\
		 Peking University, Beijing, China}

\begin{document}

\section{Introduction}

Our goal in this work to identify chiral symmetry breaking ($\chi {\rm SB}$)
below the conformal window of strongly interacting gauge theories requires the
application and testing of several methods in finite volumes.   These include
the analysis of the Goldstone spectrum in the p-regime and the spectrum of the
fermion Dirac operator with eigenvalue distributions of Random Matrix Theory
(RMT) in the $\eps$-regime, within the framework of chiral perturbation theory
($\chi {\rm PT}$). Some critical consistency checks from the theoretical
understanding of the $SU(N_f)$ rotator spectrum of the $\delta$-regime will be
also discussed. We report new results with $N_f=4,8,9,12$ flavors with
$\chi{\rm SB}$ below the conformal window for fermions in the fundamental
representation of the SU(3) color gauge group.  As $N_f$ is increased, chiral
condensate enhancement is observed when the electroweak symmetry breaking scale
$F$ is held fixed in technicolor language. We also discuss the theory inside
the conformal window. The importance of understanding finite volume,
zero momentum gauge field dynamics inside the conformal window is pointed out
and illustrated at $N_f=16$. Much of this work is an extension of our
pre-conference publication~\cite{Fodor:2009wk} where we did not report our
$N_f=12$ results.  In a forthcoming publication~\cite{JK:LongReport} more
details will be provided on the analysis and results presented here. Our work on the
running coupling is presented separately~\cite{Holland}.

It is an intriguing possibility that new physics beyond the Standard Model
might take the form of some new strongly-interacting gauge theory building on
the original technicolor
idea~\cite{Weinberg:1979bn,Susskind:1978ms,Farhi:1980xs}.  This approach has
lately been revived by new explorations of the multi-dimensional theory space
of nearly conformal gauge theories
\cite{Sannino:2009aw,Ryttov:2007sr,Dietrich:2006cm,Hong:2004td}.
Model building of a strongly interacting electroweak sector requires the
knowledge of the phase diagram of nearly conformal gauge theories as the number
of colors $N_c$, number of fermion flavors $N_f$, and the fermion
representation $R$ of the technicolor group are varied in theory space. For
fixed $N_c$ and $R$ the theory is in the chirally broken phase for low $N_f$,
and asymptotic freedom is maintained with a negative $\beta$ function. On the
other hand, if $N_f$ is large enough, the $\beta$ function is positive for all
couplings, and the theory is trivial. There is some range of
$N_f$ for which the $\beta$ function might have a non-trivial zero, an infrared
fixed point, where the theory is in fact conformal
\cite{Caswell:1974gg,Banks:1981nn}. This method has been refined by estimating
the critical value of $N_f$, above which spontaneous chiral symmetry breaking
no longer occurs \cite{Appelquist:1988yc,Cohen:1988sq,Appelquist:1996dq}. 

Interesting models require the theory to be very close to, but below, the
conformal window, with a running coupling which is almost constant over a large
energy range
~\cite{Holdom:1981rm,Yamawaki:1985zg,Appelquist:1986an,Miransky:1996pd,Kurachi:2006ej,Eichten:1979ah}.
The nonperturbative knowledge of the critical $N^{crit}_f$ separating the two
phases is essential and this has generated much interest and many new lattice
studies~\cite{Fodor:2009wk,Fodor:2008hn,Fodor:2008hm,Fodor:2009ar,Fodor:2009nh,DeGrand:2008kx,
DeGrand:2008dh,Svetitsky:2008bw,Shamir:2008pb,Appelquist:2009ty,Appelquist:2007hu,
Fleming:2008gy,DelDebbio:2008tv,DelDebbio:2008zf,DelDebbio:2008wb,Hietanen:2009az,
Hietanen:2008mr,Hietanen:2008vc,Deuzeman:2009mh,Deuzeman:2008da,Deuzeman:2008pf,
Deuzeman:2008sc,Catterall:2007yx,Catterall:2008qk,Jin:2008rc,Hasenfratz:2009ea,DeGrand:2009mt,
DeGrand:2009et,DelDebbio:2009fd,DeGrand:2009hu,Appelquist:2009ka,Jin:2009mc,Hasenfratz:2009kz,
Sinclair:2009ec}.
To provide theoretical framework for the analysis of simulation results, we
review first a series of tests expected to hold in the setting of $\chi {\rm
PT}$ in finite volume and in the infinite volume limit.

\section{Chiral symmetry breaking below the conformal window}\label{section:chpt}

We will identify in lattice simulations the chirally broken phases with
$N_f=4,8,9,12$ flavors of staggered fermions in the fundamental SU(3) color
representation using finite volume analysis. We deploy staggered fermions with
exponential (stout) smearing~\cite{Morningstar:2003gk} in the lattice action to
reduce well-known cutoff effects with taste breaking in the Goldstone
spectrum~\cite{Aoki:2005vt}.  The presence of taste breaking requires a brief
explanation of how we apply staggered $\chi {\rm PT}$ in our analysis. The
important work of Lee, Sharpe, Aubin and
Bernard~\cite{Lee:1999zxa,Aubin:2003mg,Aubin:2003uc} is closely followed in the
discussion.

\subsection{Staggered chiral perturbation theory}

Starting with $N_f=4$~\cite{Lee:1999zxa}, the spontaneous breakdown of
$SU(4)_L\times SU(4)_R $ to vector $SU(4)$ gives rise to 
15 Goldstone and pseudo-Goldstone modes,
described by fields $\phi_i$ that can be organized into an $SU(4)$ matrix
\begin{equation}
\Sigma(x) = \exp\Bigr(  i \frac{\phi }{ \sqrt{2}F}\Bigl)   \,,~~~~
\phi = \sum_{a=1}^{15} \phi_a T_a \,.
\end{equation}
$F$ is the Goldstone decay constant in the chiral limit and the normalization 
$
T_a = \left\{ \xi_\mu, i\xi_{\mu5}, i\xi_{\mu\nu}, \xi_5 \right\} 
$
is used for the flavor generators.  The leading-order chiral Lagrangian is
given by
\begin{equation}
{\cal L}_\chi^{(4)} = \frac{F^2}{4} {\sf Tr} (\partial_\mu \Sigma 
\partial_{\mu} \Sigma^{\dagger} )  - 
\frac{1}{2} B\, m_q \, F^2 {\sf Tr}( \Sigma + \Sigma^{\dagger} ) \,,
\label{eq:Lagrange}
\end{equation}
with the fundamental parameters $F$ and $B$  measured on the technicolor scale
$\Lambda_{\rm  TC}$ which replaced $\Lambda_{\rm  QCD}$ in the new theory.
Expanding the chiral Lagrangian in powers of $\phi$ one finds 15 degenerate
Goldstone pions with masses given by
\begin{equation}
M_\pi^2 = 2 B m_q \left[1 + O(m_q/\Lambda_{\rm TC}) \right] \,.
\label{eq:mpisqchiral}
\end{equation}
The leading-order term is the tree-level result while the corrections come from
loop diagrams and from higher-order terms in the chiral Lagrangian.
The addition of $a^2 {\cal L}_\chi^{(6)}$ breaks chiral symmetry and lifts the
degeneracy of the Goldstone pions. Adding correction terms to
Eq.~(\ref{eq:mpisqchiral}) yields
\begin{equation}
M_\pi^2 = C(T_a)\cdot a^2\Lambda_{\rm TC}^4 +
2 B m_q \left[1 + O(m_q/\Lambda_{\rm TC}) + O(a^2\Lambda_{\rm TC}^2) \right] 
\label{eq:Mpi2}
\end{equation}
where the representation dependent $C(T_a)$ is a constant of order unity.
Contributions proportional to $a^2$ are due to ${\cal L}_\chi^{(6)} $ and lead
to massive pseudo-Goldstone pions even in the $m_q\to 0$ chiral limit, except for the
Goldstone pion with flavor $\xi_5$ which remains massless because the $U(1)_A$ symmetry
is protected.

Lee and Sharpe observe that the part of $\CL_\chi^{(6)}$ without derivatives,
defining the potential $\CV_\chi^{(6)}$, is invariant under flavor $SO(4)$
transformations and gives rise  to the $a^2$ term in $M_\pi^2$.  Terms in
$\CL_\chi^{(6)}$ involving derivatives break $SO(4)$ further down to the
lattice symmetry group and give rise to non-leading terms proportional to $a^2
m$ and $a^4$.
The taste breaking potential is given by
\begin{eqnarray}
\hskip 0.2in - \CV_\chi^{(6)}  &=& 
 C_1 \Tr(\xi_5\Sigma \xi_5 \Sigma^\dagger)
+\, C_2\, \half 
\left[\Tr(\Sigma^2) -\Tr(\xi_5\Sigma\xi_5\Sigma) + h.c.\right] \nonumber \\
&+&\, C_3\, \half \sum_\nu \left[\Tr(\xi_\nu\Sigma\xi_\nu \Sigma) + h.c.\right] 
+\, C_4\, \half \sum_\nu 
\left[\Tr(\xi_{\nu5}\Sigma\xi_{5\nu} \Sigma) + h.c.\right]\nonumber \\
&+&\, C_5\, \half \sum_\nu \left[\Tr(\xi_\nu\Sigma \xi_\nu \Sigma^\dagger) 
         - \Tr(\xi_{\nu5}\Sigma \xi_{5\nu} \Sigma^\dagger) \right]
+\, C_6 \sum_{\mu<\nu} 
\Tr(\xi_{\mu\nu}\Sigma \xi_{\nu\mu} \Sigma^\dagger) \,.
\label{eq:massterms}
\end{eqnarray}
The six unknown coefficients $C_i$ are all of size $\Lambda_{\rm TC}^6$.

In the continuum, the Goldstone pions form a 15-plet of flavor $SU(4)$ and are
degenerate. On the lattice, states are classified by the symmetries of the
transfer matrix, and the pseudo-Goldstone  pions fall into 7 irreducible
representations: four 3-dimensional representations with flavors $\xi_i$,
$\xi_{i5}$, $\xi_{ij}$ and $\xi_{i4}$, and three 1-dimensional representations
with flavors $\xi_4$, $\xi_{45}$ and $\xi_5$.

Close to both the chiral and continuum limits, the pseudo-Goldstone masses are given by
\begin{equation}
M_\pi(T_a)^2 = 2 B m_q + a^2 \Delta(T_a) + O(a^2 m_q) + O(a^4) \,,
\label{eq:pimassform}
\end{equation}
with $\Delta(T_a) \sim \Lambda_{\rm TC}^4$ arising from $\CV_\chi^{(6)}$.
Since $\CV_\chi^{(6)}$ respects flavor $SO(4)$, the 15 Goldstone particles fall
into $SO(4)$ representations:
\begin{eqnarray}
  &&\Delta(\xi_5) = 0, \hskip 0.1in  \Delta(\xi_\mu) = {8 \over F^2} 
(C_1+C_2+C_3+3 C_4+C_5+3 C_6)  ,\nonumber \\ 
&& \Delta(\xi_{\mu5}) ={8 \over F^2} 
(C_1+C_2+3 C_3+ C_4-C_5+3 C_6) , 
\hskip 0.1in \Delta(\xi_{\mu\nu}) = {8 \over F^2} (2C_3+2 C_4+4 C_6) .
\label{eq:ximu5}
\end{eqnarray}
In the chiral limit at finite lattice spacing, the lattice irreducible
representations with flavors $\xi_i$ and $\xi_4$ are degenerate,  those with
flavors $\xi_{i5}$ and $\xi_{45}$, and those with flavors $\xi_{ij}$ and
$\xi_{i4}$ are degenerate as well.  No predictions can be made for the
ordering, splittings, or even the {\em signs} of the mass shifts.  Our
simulations indicate that they are all positive with the exponentially smeared
staggered action we use, making the existence of an Aoki
phase~\cite{Lee:1999zxa} unlikely.  The method of~\cite{Lee:1999zxa} has been
generalized to the $N_f > 4$ case~\cite{Aubin:2003mg,Aubin:2003uc} which we
adopted in our calculations with help from Bernard and Sharpe. The procedure
cannot be reviewed here but it will be used in the interpretation of our
$N_f=8$ simulations.

\subsection{Finite volume analysis in the p-regime}

Three different regimes can be selected in simulations to identify the chirally
broken phase from finite volume spectra and correlators.  For a  lattice size
$L_s^3\times L_t$ in euclidean space and in the limit $L_t \gg  L_s$, the
conditions $F_{\pi}L_s > 1$ and $M_{\pi}L_s > 1$ select the the p-regime, in
analogy with low momentum counting~\cite{Gasser:1987zq,Hansen:1990yg}. 

For arbitrary $N_f$, in the continuum and in infinite volume, the one-loop
chiral corrections  to $M_\pi$ and $F_\pi$ of the degenerate Goldstone pions
are given by
\begin{equation}
      ~~~~  M^2_\pi = M^2  \biggl [1-\frac{M^2}{8\pi^2N_fF^2}ln\biggl (\frac{\Lambda_3}{M}\biggr ) \biggr ] ,
\label{eq:Mpi}                         
\end{equation}
\begin{equation}
                 F_\pi = F  \biggl [1+\frac{N_fM^2}{16\pi^2F^2}ln\biggl (\frac{\Lambda_4}{M}\biggr ) \biggr ] ,
\label{eq:Fpi}
\end{equation}
where $M^2=2B\cdot m_q$ and $F,B,\Lambda_3,\Lambda_4$ are four fundamental
parameters of the chiral Lagrangian, and the small quark mass $m_q$ explicitly
breaks the symmetry~\cite{Gasser:1983yg}. The chiral parameters $F,B$ appear in
the leading part of the Lagrangian in Eq.~(\ref{eq:Lagrange}), while
$\Lambda_3,\Lambda_4$ enter in next order. There is the well-known GMOR
relation $\Sigma_{cond}=BF^2$ in the $m_q \rightarrow 0$ limit for the  chiral
condensate per unit flavor~\cite{GellMann:1968rz}.  It is important to note
that the one-loop correction to the pion coupling constant $F_{\pi}$ is
enhanced by a factor $N_f^2$ compared to $M_{\pi}^2$.  The chiral expansion for
large $N_f$ will break down for $F_{\pi}$ much faster for a given
$M_{\pi}/F_{\pi}$ ratio. The NNLO terms have been recently 
calculated~\cite{Bijnens:2009qm} showing potentially dangerous $~ N^2_f$
corrections to Eqs.~(\ref{eq:Mpi},\ref{eq:Fpi}).

The finite volume corrections to $M_\pi$ and $F_\pi$ are given in the p-regime
by
\begin{equation}
      M_\pi(L_s,\eta) = M_\pi  \biggl [1+\frac{1}{2N_f}\frac{M^2}{16\pi^2F^2}\cdot\tilde g_1(\lambda,\eta) \biggr ] ,
\label{eq:MpiL}
\end{equation}
\begin{equation}
     F_\pi (L_s,\eta)= F_\pi \biggl [1-\frac{N_f}{2}\frac{M^2}{16\pi^2F^2} \cdot\tilde g_1(\lambda,\eta) \biggr ] ,
\label{eq:FpiL}
\end{equation}
where $\tilde g_1(\lambda,\eta)$ describes the finite volume corrections with
$\lambda=M\cdot L_s$ and aspect ratio $\eta=L_t/L_s$.  The form of $\tilde
g_1(\lambda,\eta)$ is a complicated infinite sum which contains Bessel
functions and requires numerical evaluation~\cite{Hansen:1990yg}.
Eqs.~(\ref{eq:Mpi}-\ref{eq:FpiL}) provide the foundation of the p-regime fits
in our simulations.

\begin{figure}[ht!]
\begin{center}
\includegraphics[width=8cm,angle=0]{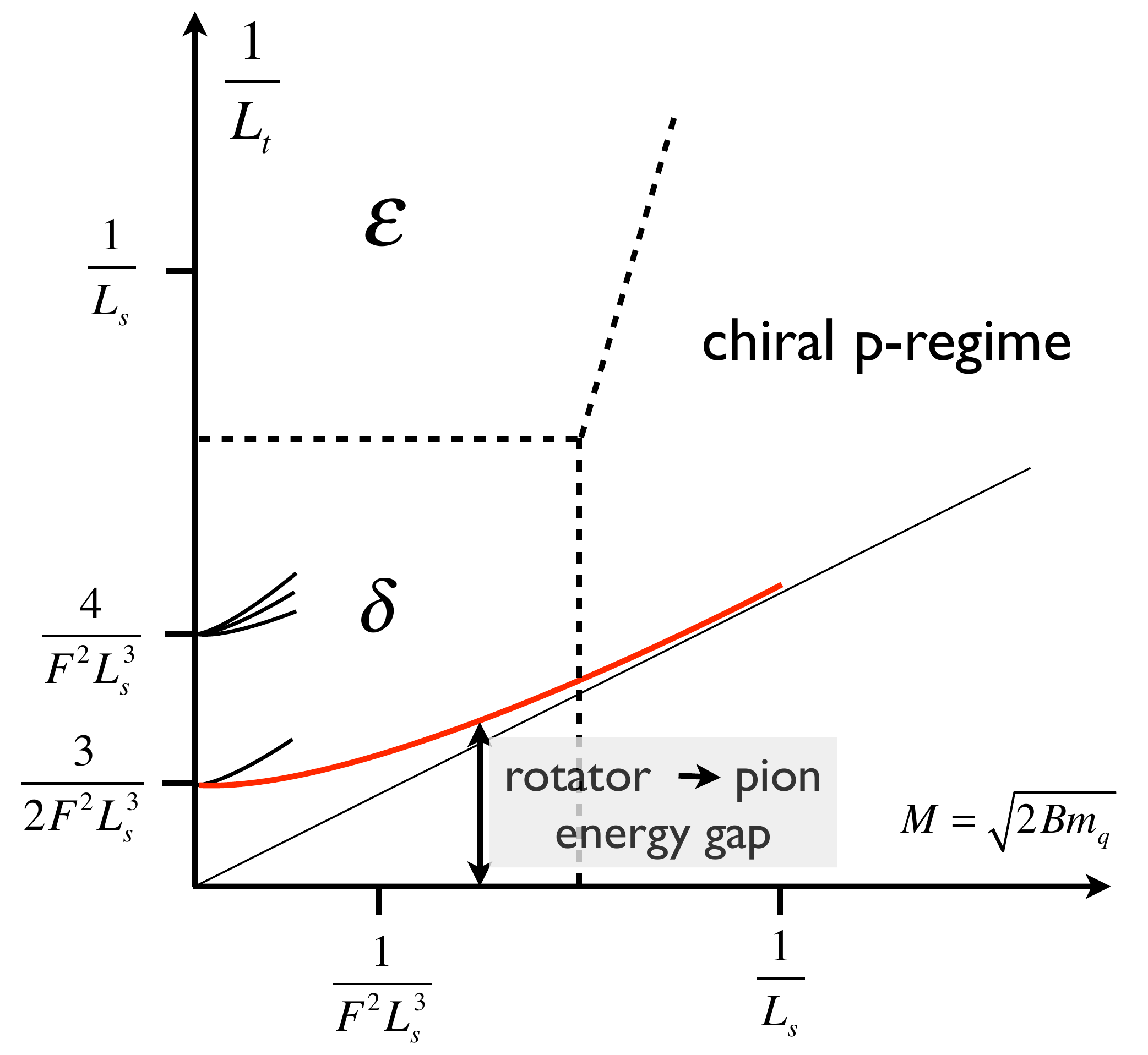} 
\end{center}
\vskip -0.1in
\caption{Schematic plot of the regions in which the three low energy chiral
expansions are valid.  The vertical axis shows the finite temperature scale
(euclidean time in the path integral) which probes the rotator dynamics of the
$\delta$-regime and the $\eps$-regime. The first two low lying rotator
levels are also shown on the vertical axis for the simple case of  $N_f=2$. The
fourfold degenerate lowest rotator excitation at $m_q=0$ will split into an
isotriplet state (lowest energy level), which evolves into the p-regime pion as
$m_q$ increases, and into an isosinglet state representing a multi-pion state
in the p-regime. Higher rotator excitations have similar interpretations.}   
\label{fig:landscape}
\vskip -0.1in
\end{figure}

\subsection{$\delta$-regime and $\eps$-regime}

At fixed $L_s$ and in cylindrical geometry $L_t/L_s \gg 1$,  a crossover occurs
from the p-regime to the $\delta$-regime when $m_q \rightarrow 0$, as shown in
Fig.~\ref{fig:landscape}.  The dynamics is dominated by the rotator states of
the chiral condensate in this limit~\cite{Leutwyler:1987ak} which is
characterized by the conditions $FL_s > 1$ and $ML_s \ll 1$.  The densely
spaced rotator spectrum scales with gaps of  the order $\sim 1/F^2L_s^3$, and
at $m_q=0$ the chiral symmetry is apparently  restored. However, the rotator
spectrum, even at $m_q=0$ in the finite volume, will  signal that the infinite
system is in the chirally broken phase for the particular parameter set of the
Lagrangian. This is often misunderstood in the interpretation of lattice
simulations.  Measuring finite energy levels with pion quantum numbers at fixed
$L_s$ in the  $m_q \rightarrow 0$ limit is not a signal for chiral symmetry
restoration of the infinite system~\cite{Deuzeman:2009mh}.
\begin{figure}[h!]
\begin{center}
\includegraphics[width=9cm,angle=0]{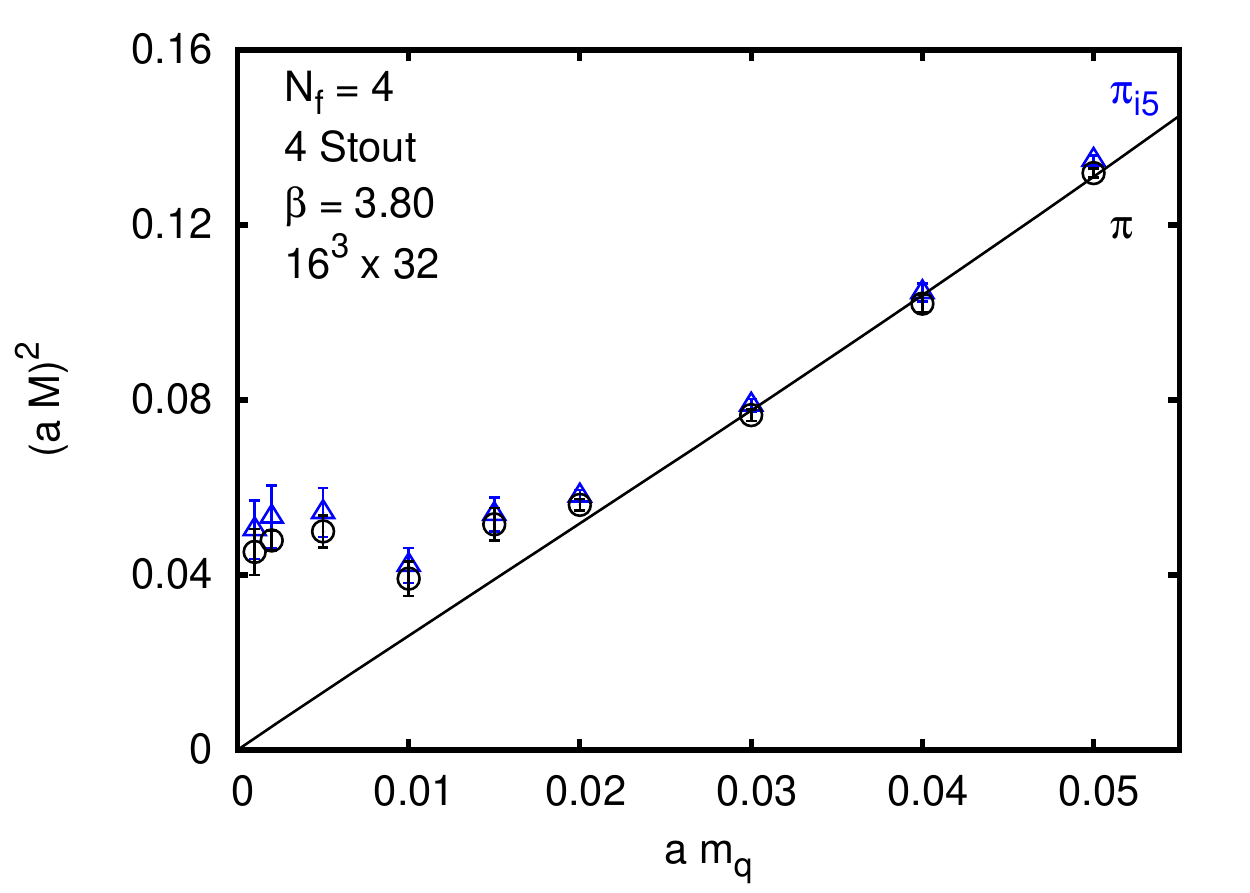}
\end{center}
\vskip -0.2in
\caption{The crossover from the p-regime to the $\delta$-regime is shown for
the $\pi$ and $\pi_{i5}$ states at $N_f=4$. }   
\label{fig:delta}
\vskip -0.05in
\end{figure}

If $L_t \sim L_s$ under the conditions $FL_s > 1$ and $ML_s \ll 1$, the system
will be driven into the $\eps$-regime which can be viewed as the high
temperature limit of the $\delta$-regime quantum rotator.  Although the
$\delta$-regime and $\eps$-regime have an overlapping region, there is an
important difference in their dynamics. In the $\delta$-regime of the quantum
rotator, the mode of the pion field $U(x)$ with zero spatial momentum dominates
with time-dependent quantum dynamics.  The $\eps$-regime is dominated by
the four-dimensional zero momentum  mode of the  chiral Lagrangian. 

We report simulation results of all three regimes in the chirally broken phase
of the technicolor models we investigate. The analysis of the three regimes
complement each other and provide cross-checks for the correct identification
of the phases. First, we will probe Eqs.~(\ref{eq:Mpi}-\ref{eq:FpiL}) in the
p-regime, and follow with the study of Dirac spectra and RMT eigenvalue
distributions in the $\eps$-regime. The spectrum in the $\delta$-regime is
used as a signal to monitor p-regime spectra as $m_q$ decreases.
Fig.~\ref{fig:delta} is an illustrative example of this crossover in our
simulations. It is important to note that the energy levels in the chiral limit
do not match the rotator spectrum at the small $F\cdot L_s$ values of the
simulations. This squeezing with $F\cdot L_s$ not large enough for undistorted,
finite volume, chiral behavior in the p-regime, $\eps$-regime, and
$\delta$-regime, will be further discussed in our p-regime simulations
presented next. We will also describe some methods to put this squeezing into a more
quantitative context.

\section {Goldstone spectra and $\chi{\rm SB}$ from simulations at $ \bf  N_f=4 $  in the p-regime}

\begin{figure}[ht!]
\begin{center}
\begin{tabular}{cc}
\includegraphics[height=5.3cm]{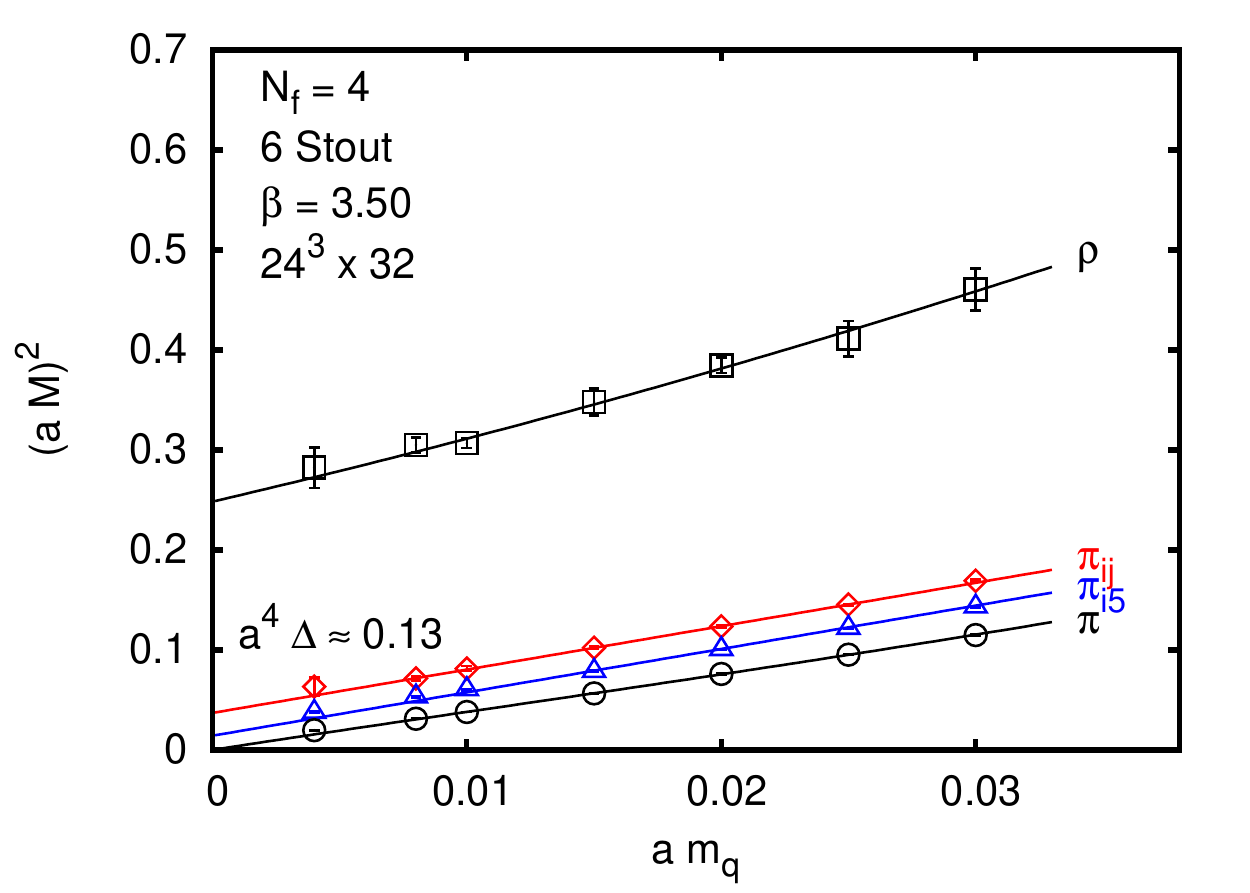}&
\includegraphics[height=5.3cm]{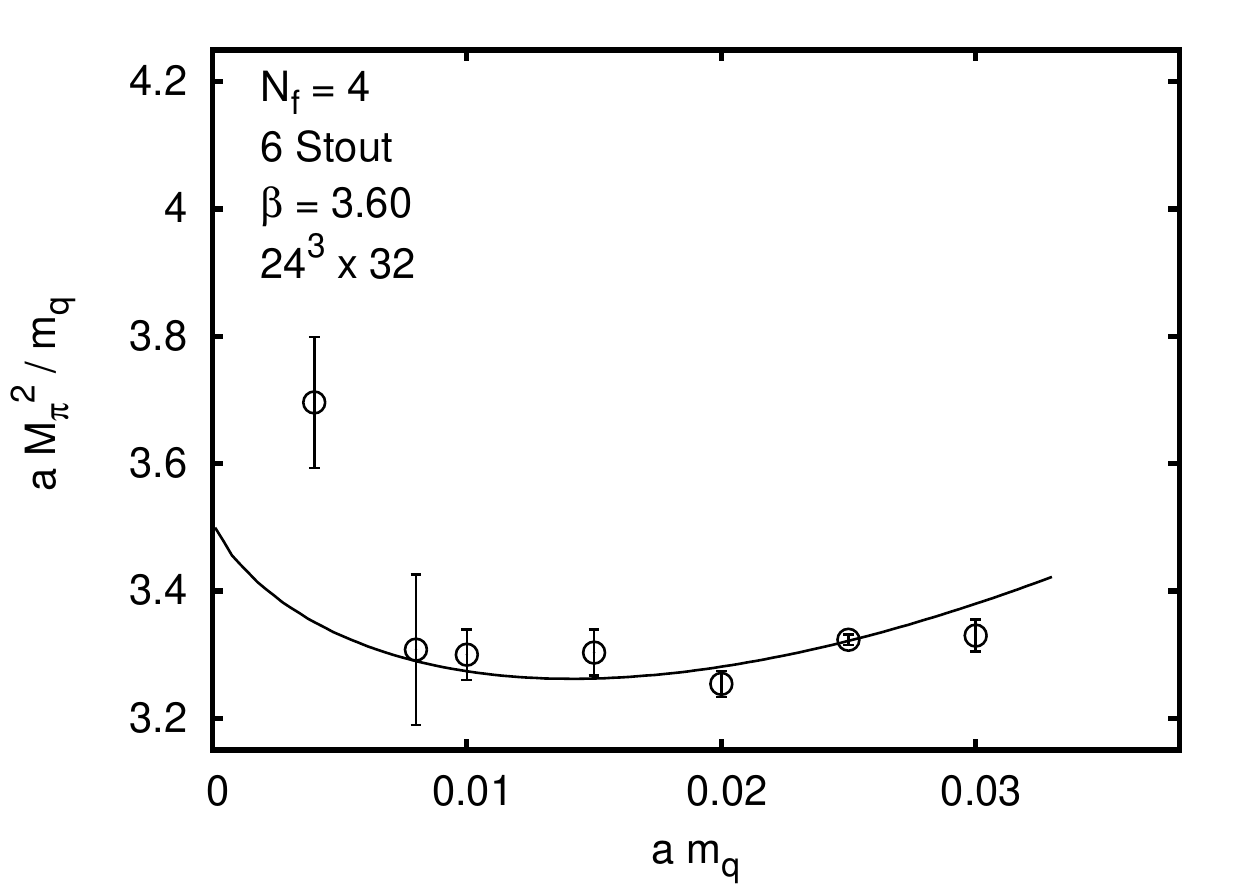}\\
\includegraphics[height=5.3cm]{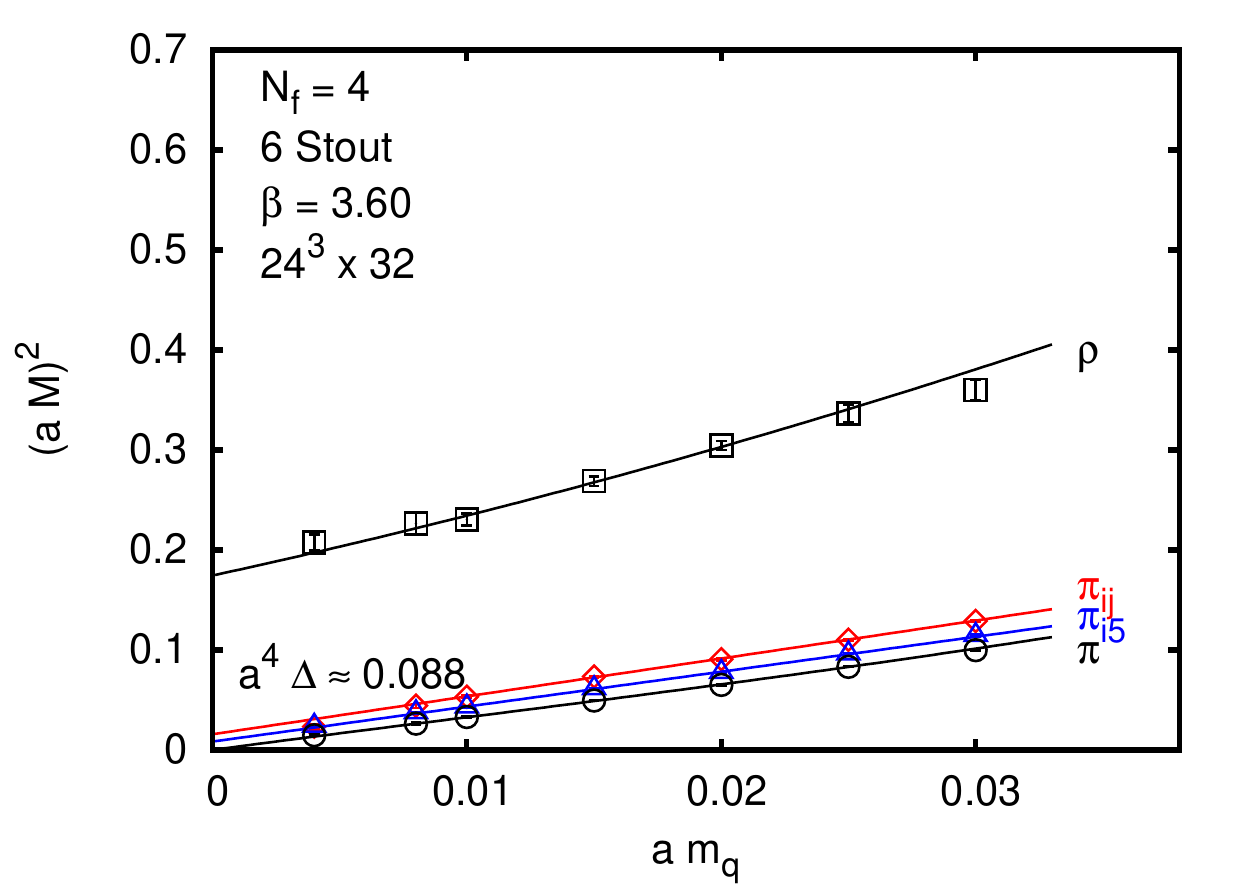}&
\includegraphics[height=5.3cm]{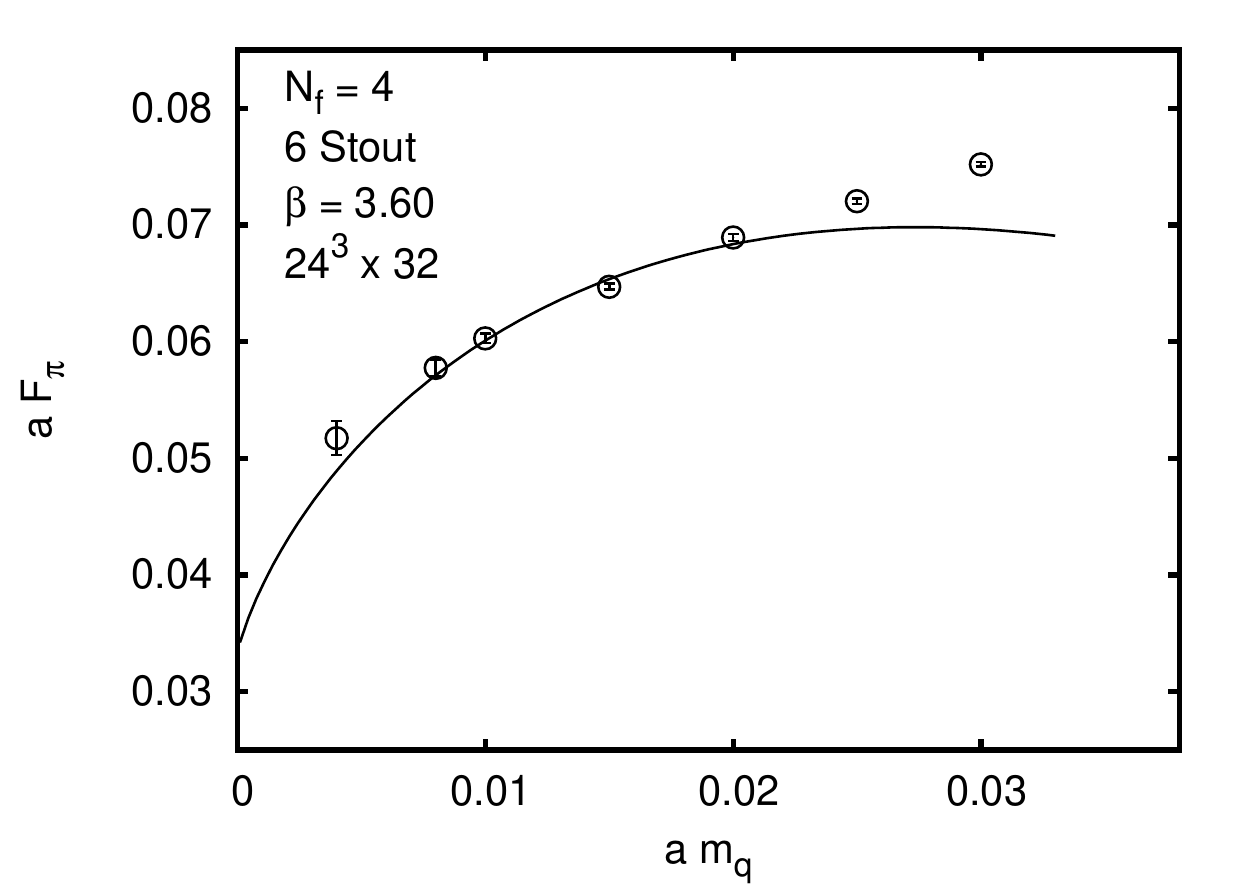}\\
\includegraphics[height=5.3cm]{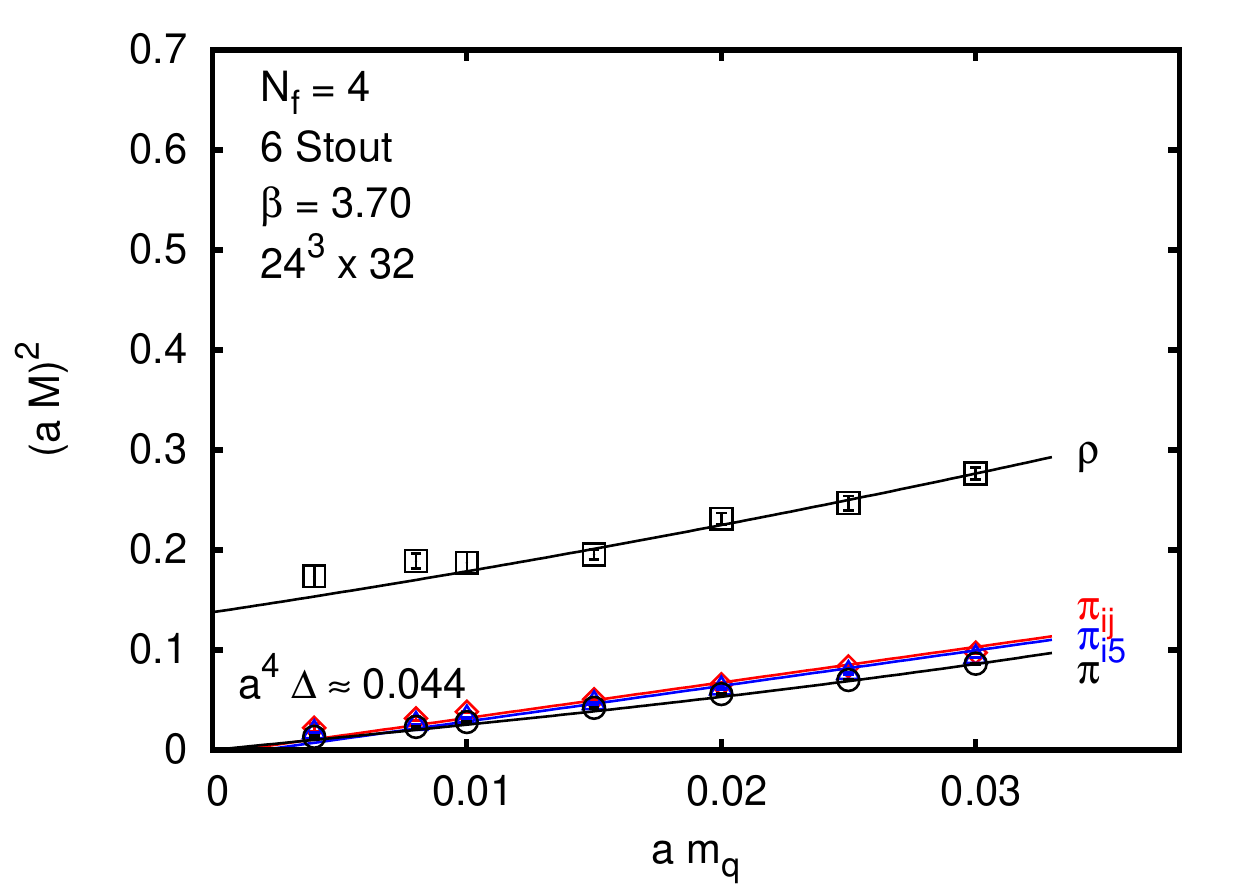}&
\includegraphics[height=5.3cm]{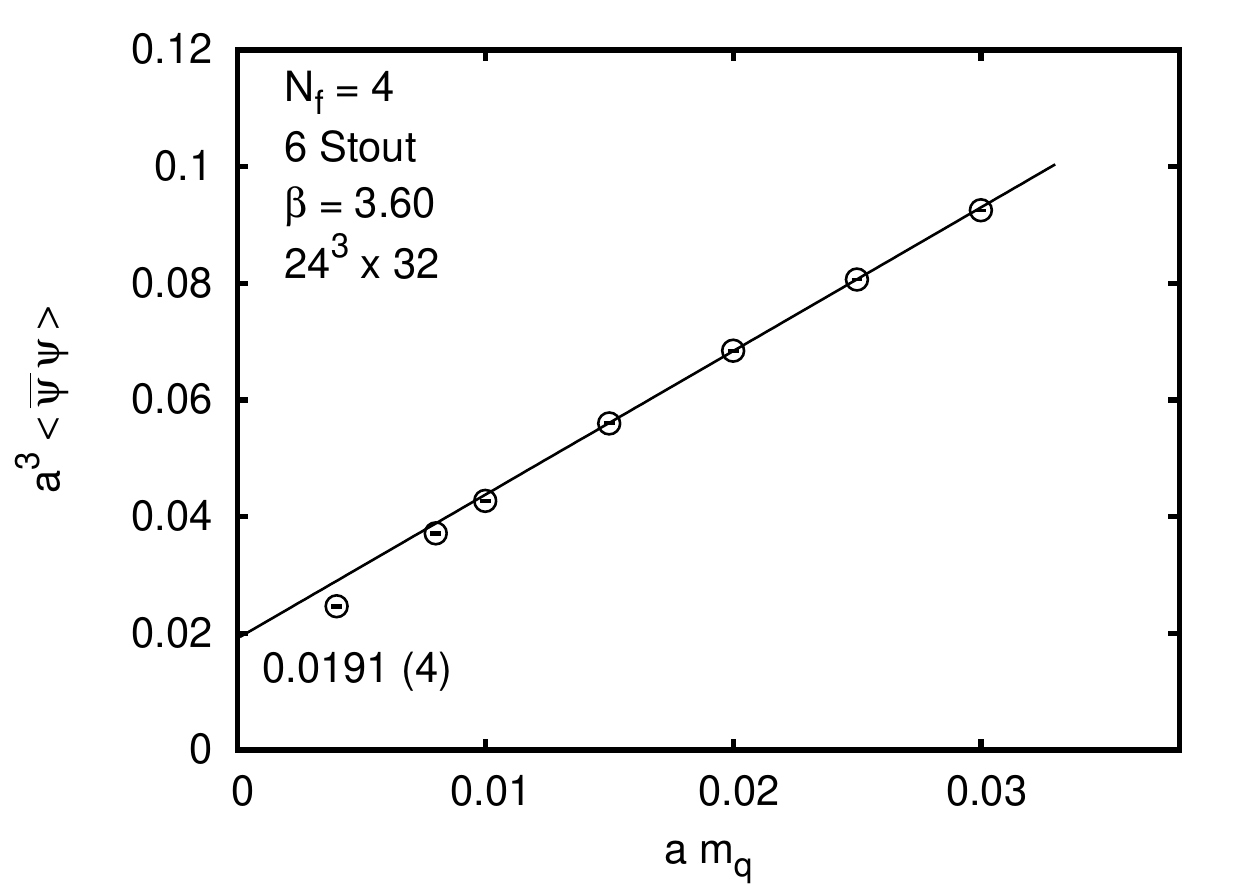}
\end{tabular}
\caption{The Goldstone spectrum and chiral fits are shown for simulations with
lattice size $24^3\times 32$. The left column shows the pseudo-Goldstone spectrum with
decreasing taste breaking as the gauge coupling is varied from $\beta=3.5$ to
$\beta=3.7$.  The middle value at $\beta=3.6$ was chosen for chiral fits which
are shown in the right column. The top right figure with fitting range $a\cdot
m_q=0.008-0.025$ shows the NLO chiral fit to $M^2_\pi/m_q$ which approaches
$2B$ in the chiral limit. Data points below $m_q=0.008$ are not in the chiral
p-regime and not used in the fitting procedure.The middle figure on the right
is the NLO chiral fit to $F_\pi$ in the range $a\cdot m_q=0.008-0.02$. The
bottom right figure is the linear fit to the chiral condensate with fitting
range $a\cdot m_q=0.015-0.025$. The physical fit parameters
$B,F,\Lambda_3,\Lambda_4$ are discussed in the text.}   
\end{center}
\label{fig:Nf4}
\vskip -0.2in
\end{figure}

In this section we describe in some detail  the methods we use for successfully
testing chiral symmetry breaking.  Our tests in the p-regime have two major
components. The primary test is to identify the pseudo-Goldstone spectrum of the
staggered formulation with evidence for recovery from taste symmetry breaking
close to the continuum limit. The secondary test is to probe chiral loop
corrections to the tree-level behavior of $M^2_\pi$ and $F_\pi$ as the fermion
mass $a\cdot m_q$ is varied at fixed gauge coupling $\beta$.  The evidence we
find for chiral symmetry breaking at $N_f=4,8,9,12$ is common to all
flavors.  Limitations and ambiguities identified at $N_f=4$ for future
improvements are expected to be more pronounced with increasing $N_f$.  Results
for each flavor we have simulated in the p-regime are presented in separate
sections beginning here with general discussion and $N_f=4$ results. 

We have used the tree-level Symanzik-improved gauge action for all simulations.
The conventional $\beta=6/g^2$ lattice gauge coupling is defined as the overall
factor in front of the three well-known terms of the lattice action.  The link
variables in the staggered fermion matrix were exponentially smeared with  six
stout steps at $N_f=4$  and the RHMC algorithm was deployed in all runs. The
results shown in Fig.~\ref{fig:Nf4} are from the p-regime of $\chi{\rm SB}$
with the conditions  $M_\pi\cdot L_s \gg 1$ and $F_\pi\cdot L_s \sim 1$  when
the chiral condensate begins to follow the expected behavior of infinite-volume
chiral perturbation theory from Eqs.~(\ref{eq:Mpi},\ref{eq:Fpi}) in
next-to-leading order (NLO) with calculable finite volume corrections from
Eqs.~(\ref{eq:MpiL},\ref{eq:FpiL}) which are negligible at $L_s=24$.  We have
empirical evidence that the  $M_\pi$  and $F_\pi$ data points are free of
finite volume corrections in practically the entire fitting range of the
fermion masses we use at $L_s=24$ so that the negligible corrections from
Eqs.~(\ref{eq:MpiL},\ref{eq:FpiL}) can be ignored.

Within some finite volume limitations, which we will address, the $N_f=4$
simulations work in the p-regime as expected.  The left column of
Fig.~\ref{fig:Nf4} shows that the pseudo-Goldstone spectrum clearly remains separated
from the hadronic scale of the $\rho$-meson as $\beta$ is varied.  Moving
towards the continuum limit with increasing $\beta$, we see the split pseudo-Goldstone
spectrum collapsing into the degenerate continuum  pion spectrum. The true
Goldstone pion whose mass will vanish in the $a\cdot m_q=0$ limit at fixed lattice
spacing and two additional split states with small residual masses at $a\cdot
m_q = 0$  are shown to illustrate the trend.  $a^4\Delta$ is the measure of the
small taste breaking in quadratic mass splitting as measured in lattice units.
The origin of the splittings and the quantum numbers were discussed in Section
\ref{section:chpt} as shown in Eq.~(\ref{eq:ximu5}).  The
spectrum is approximately parallel as the bare fermion mass $a\cdot m_q$ is
varied at fixed lattice spacing and the gaps appear to be equally spaced to a
good approximation, consistent with earlier observations in QCD where the $C_4$
term seems to dominate staggered taste breaking for two light flavors with
equally spaced pseudo-Goldstone levels~\cite{Lee:1999zxa}.  We selected $\beta=3.6$
for testing $\chi {\rm PT}$ of finite volume Goldstone dynamics in the
p-regime.  This choice with small taste breaking is close to the continuum
limit without excessive squeeze on the important product $F\cdot L_s$  which in
an ideal simulation of the p-regime should be large ($F$ is the chiral limit of
$F_\pi$ as $a\cdot m_q \rightarrow 0$ at fixed lattice spacing).

The simultaneous chiral fit of  $M^2_\pi/m_q$ and $F_\pi$ based on
Eqs.~(\ref{eq:Mpi}-\ref{eq:FpiL}) is shown in Fig.~\ref{fig:Nf4} where chiral
loops correct the tree-level values of  $M^2_\pi/m_q=2B$ and $F_\pi$.  In the
fitting range $a\cdot m_q=0.008 - 0.025$ applied to $M^2_\pi/m_q$ we observe
small corrections to the tree-level value of $2B$ which keeps the fit well
within the range of one-loop $\chi {\rm PT}$.  In the fitting range $a\cdot
m_q=0.008 - 0.02$ the $F_\pi$ data are about a factor of two larger than $F$
which indicates how the one-loop fit is being pushed to its limits.  Without
loop correction $F_\pi$ would not change from its fitted value of  
$a\cdot F=0.033(4)$ 
in the chiral limit at fixed lattice spacing.  The fitted value of $B$ is 
$a\cdot B=1.76(7)$
in lattice units and 
$M_\rho/F=13(1)$ 
in the chiral limit (the linear fit of $M_\rho={\rm c + d\cdot m_q}$ is used at
all $N_f$ values to determine $M_\rho({\rm m_q=0})$). The fitted value of 
$B/F = 53(6)$ 
indicates significant enhancement of the chiral condensate from its $N_f=2$
value~\cite{Appelquist:2009ka,Noaki:2008iy}. In our simultaneous fits we get
$\Lambda_3=0.37(5)$ 
and 
$\Lambda_4=0.51(1)$ 
which set the chiral couplings in the NLO chiral Lagrangian.

The chiral condensate $\langle \bar{\psi}\psi\rangle$ summed over all flavors
is dominated by the linear term in $m_q$ from UV contributions.  The linear fit
gives 
$\langle \bar{\psi}\psi\rangle=0.0191(4)$ 
in the chiral limit which differs from the GMOR relation of  $\langle
\bar{\psi}\psi\rangle=4F^2B$  by about a factor of two with 
$4F^2B=0.008(2)$ 
fitted. There are several sources of this disagreement. The chiral log in
$\langle \bar{\psi}\psi\rangle$ will bring further down the true fitted value
in the chiral limit. Our volumes are not large enough yet to attempt a sensible
chiral log fit to the condensate at small $a\cdot m_q$ values. Finite volume
squeezing effects distort the consistency of the results in our limited range
of simulation volumes.  The choice of fitting method to
Eqs.~(\ref{eq:Mpi},\ref{eq:Fpi}) can also have some effect on the results.  On
the right-hand sides of the equations, the variable pair $(M,~F)$ in the chiral
logs can be replaced with the pair $(M_\pi,~F_\pi)$ which is equivalent to a
partial resummation~\cite{Noaki:2008iy}.  This will be reported in our more
detailed forthcoming journal publication~\cite{JK:LongReport}.

Finite volume limitations when measured in $F$ units have the biggest effect on
our chiral analysis.  The value $F\cdot L_s \approx 0.8$ is not expected to
protect against significant finite-volume squeezing effects for even the
largest spatial size $L_s=24$ used in the chiral fits at $N_f=4$.  Larger than
optimal NLO corrections in the chiral fits of $F_\pi$ and  finite-volume
squeezing effects  are closely related concerns. Simulations on larger lattices
would increase $F\cdot L_s$ and allow us to drop back in $a\cdot m_q$ into a
more comfortable range with smaller NLO chiral corrections for $F_\pi$.

Finite volume corrections to the rotator spectrum in the $\delta$-regime set
some quantitative measure of squeezing effects on the chiral analysis. The
connection is made by observing that the pion spectrum in the p-regime can be
viewed at fixed spatial volume $L^3_s$ as the adiabatic evolution from the
energy levels of the rotator spectrum of the $\delta$-regime as illustrated
schematically in Fig.~\ref{fig:delta} for the lowest $N_f=2$ rotator levels.  The
rotator spectrum for $N_f=2$ is given by $E_l = \frac{1}{2\Theta}l(l+2)$, with
$l=0,1,2, ...$, where the moment of inertia is calculated in
NLO~\cite{Hasenfratz:1993vf,Hasenfratz:2009mp} as $\Theta =
F^2L^3_s(1+\frac{C(N_f=2)}{F^2L^2_s} + O(\frac{1}{F^4\cdot L^4_s}) )$.  The
value of $C(N_f=2)$ is known to be 0.45 and is expected to grow with $N_f$.
At $F\cdot L_s \approx 0.8$ the correction is 70\% for $N_f=2$ and 
probably considerably larger for $N_f=4$. The leading-order rotator gap for
arbitrary $N_f$ is given by  $E_1-E_0= \frac{N^2_f - 1}{N_fF^2L^3_s}$ but the
coefficient $C(N_f)$ is an important missing piece in the analysis.  Were we to
continue the p-regime Goldstone spectrum at $N_f=4$, $L_s=24$, and $\beta=3.6$ to
the $\delta$-regime adiabatically, the small value of $F\cdot L_s$ would not
allow us to get a reliable estimate of $F$ based on the chiral rotator spectrum
with the collapse of the adiabatic approximation. This is a quantitative
warning sign of the need for considerably larger spatial volume for robust
p-regime results to determine $F$ in the chiral fitting procedure.  In fact, we
are going to fit $M^2_\pi/m_q$  in the chiral analysis for $N_f=8,9,12$ with
better controlled NLO loop corrections, but $F_\pi$ will not be fitted.  For
larger $N_f$, a reliable simultaneous fit requires substantially larger volumes
than are realistic with our current resources. 

In summary, the $N_f=4$ system passed both tests in the chirally broken phase
and shows significant enhancement of the chiral condensate when measured in
units of the electroweak symmetry breaking scale set by $F$. This is a relevant
effect to monitor for fermion mass generation in extended technicolor
applications as we begin to approach the conformal
window~\cite{Appelquist:2009ka}.

\section {Goldstone spectrum and $\chi{\rm SB}$ from simulations at $\bf  N_f=8 $  in the p-regime}

\begin{figure}[ht!]
\begin{center}
\begin{tabular}{cc}
\includegraphics[height=5.3cm]{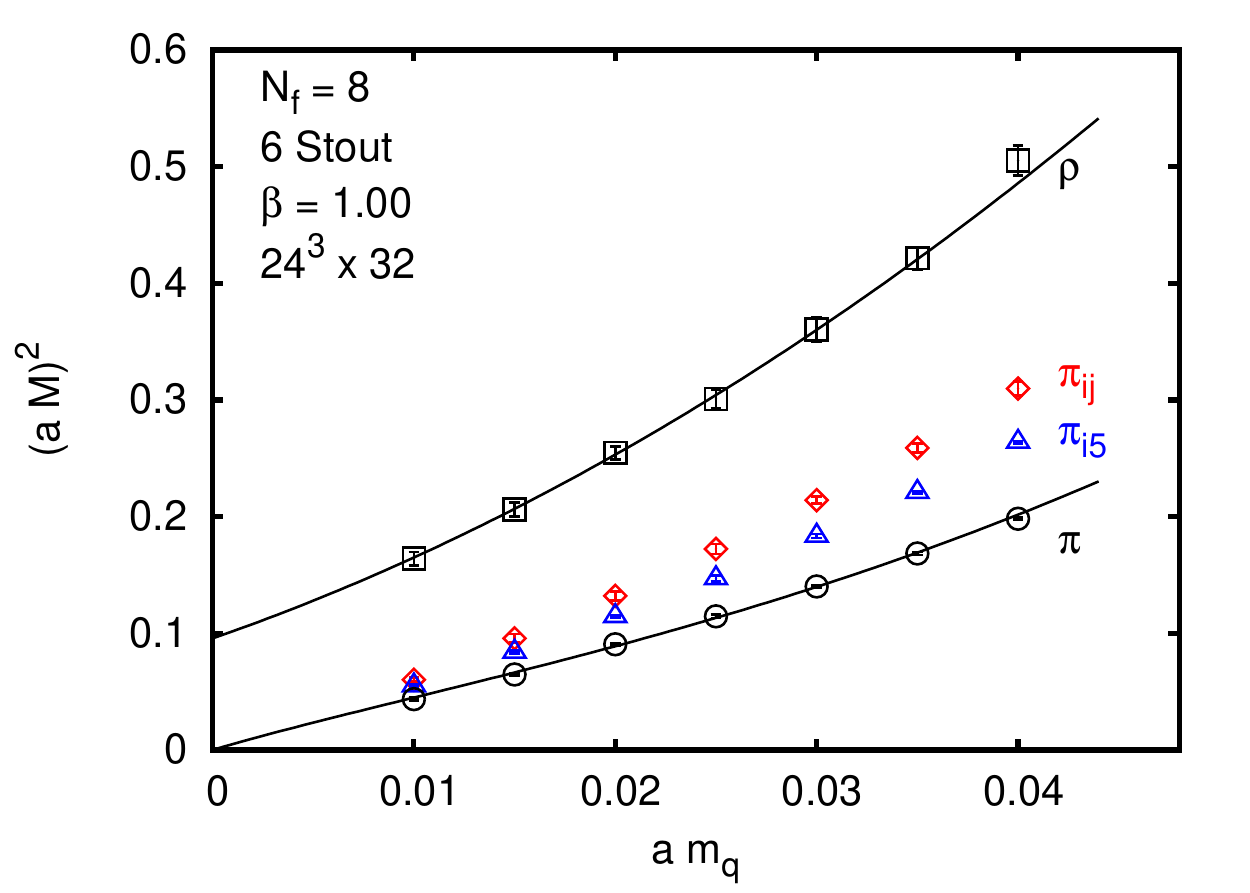}&
\includegraphics[height=5.3cm]{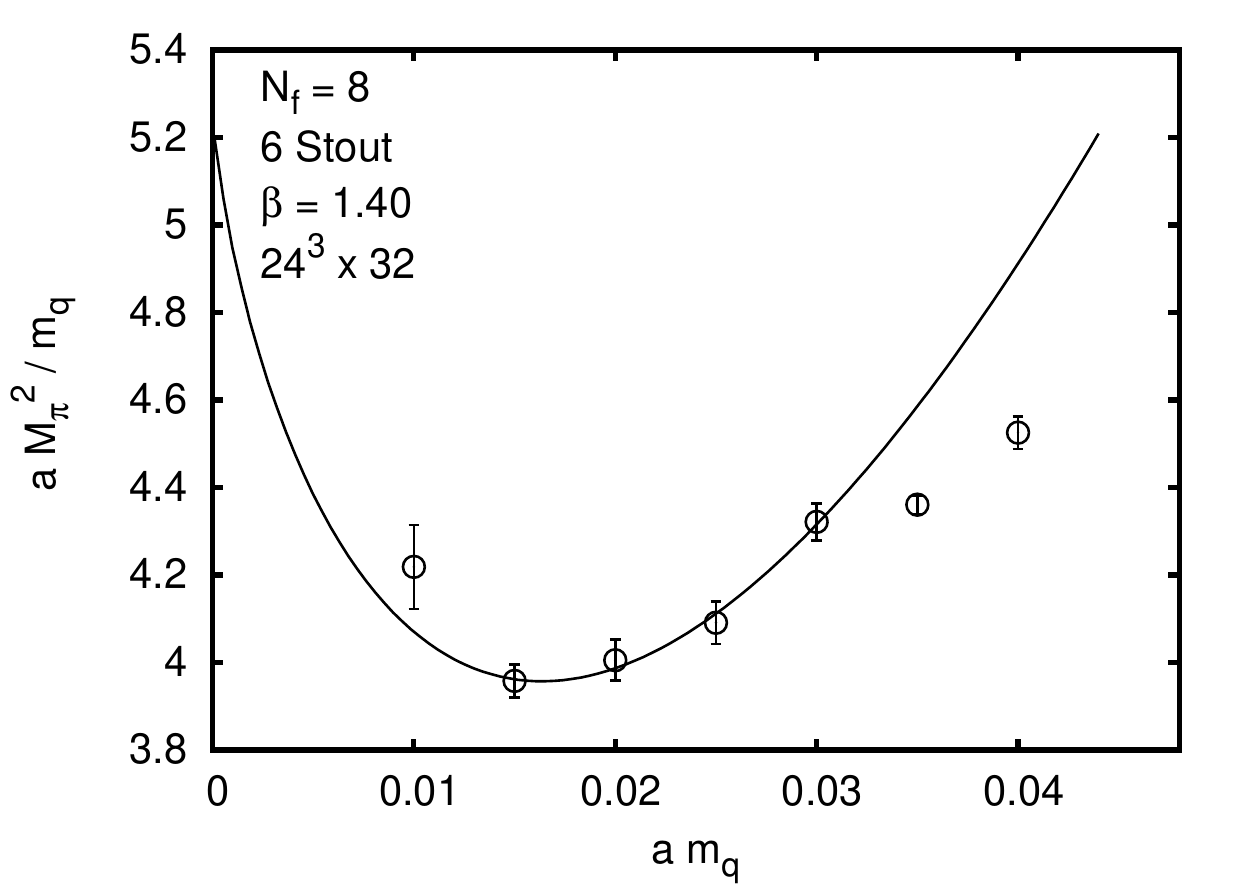}\\
\includegraphics[height=5.3cm]{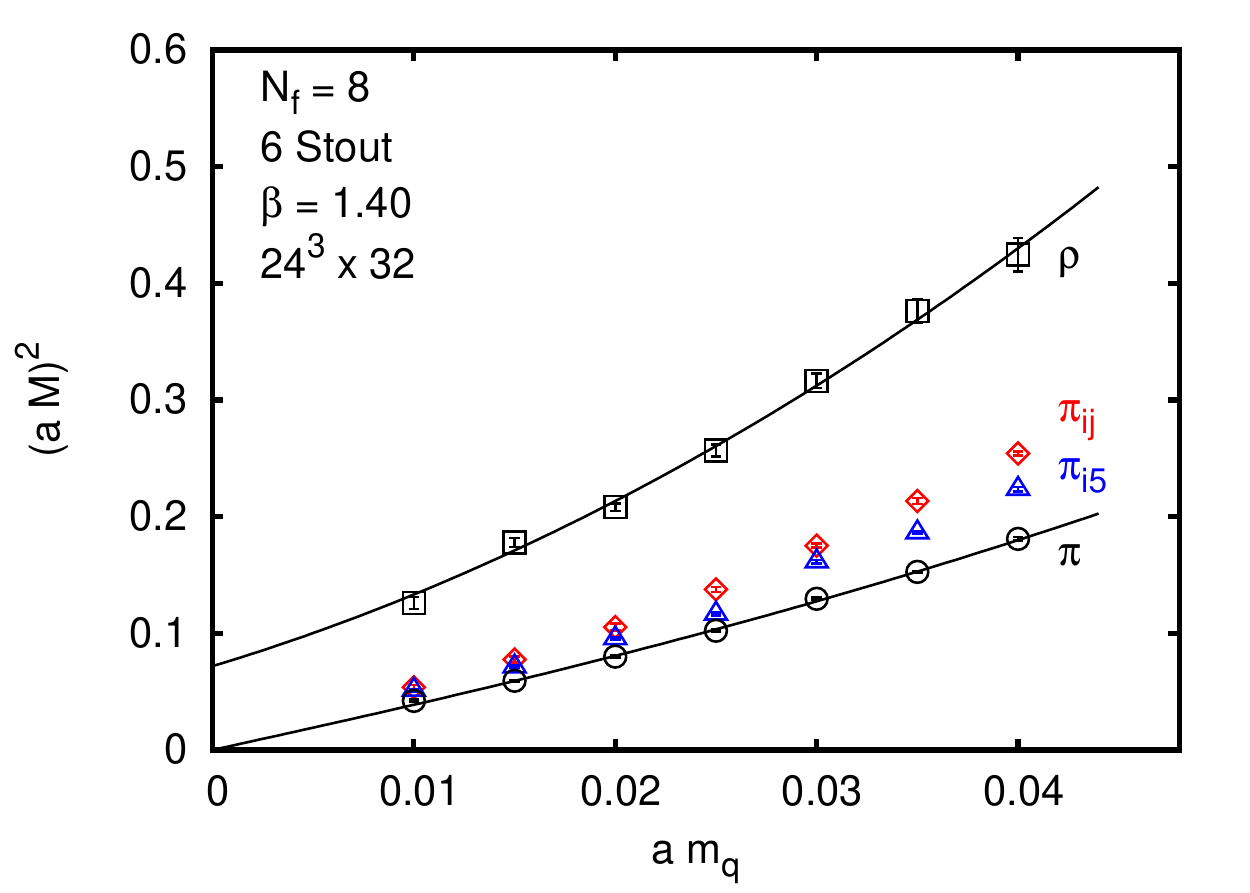}&
\includegraphics[height=5.3cm]{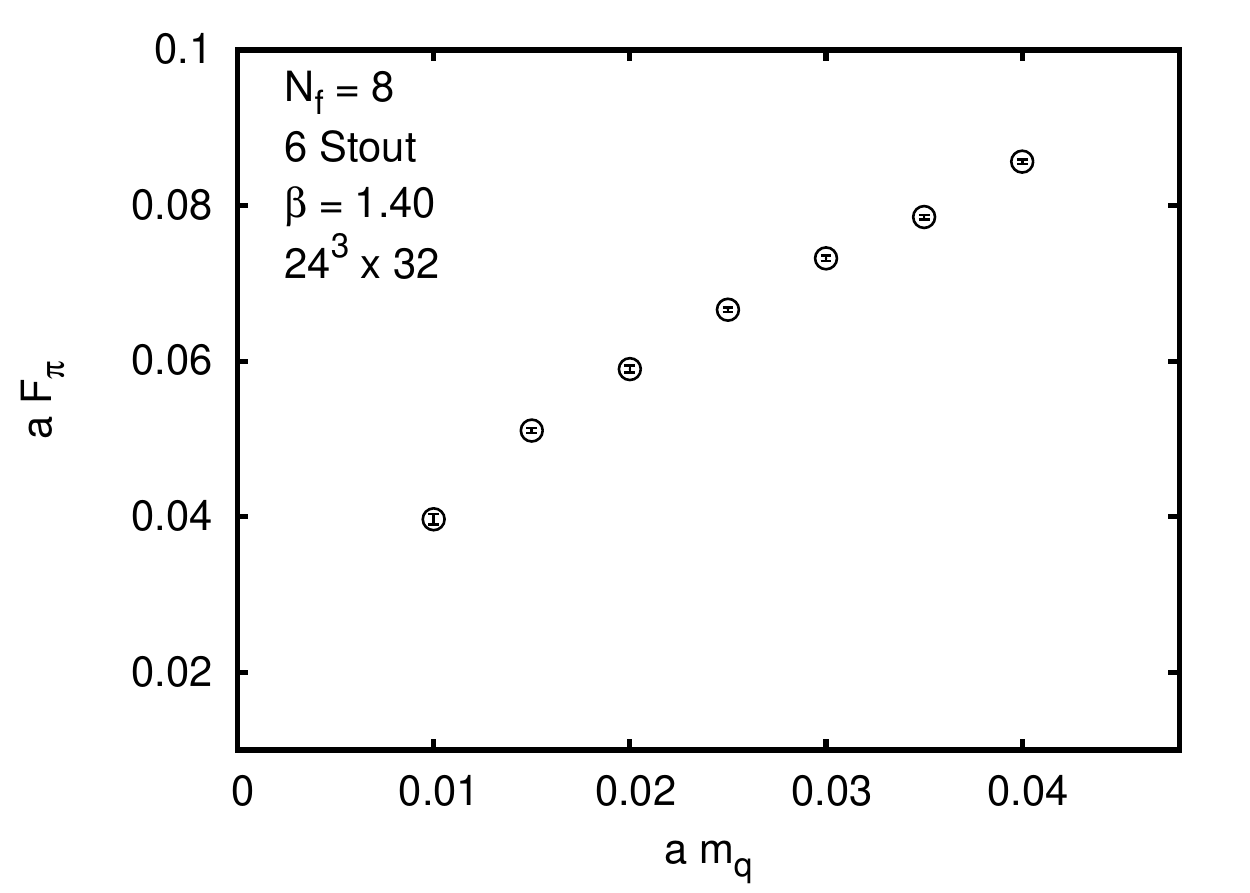}\\
\includegraphics[height=5.3cm]{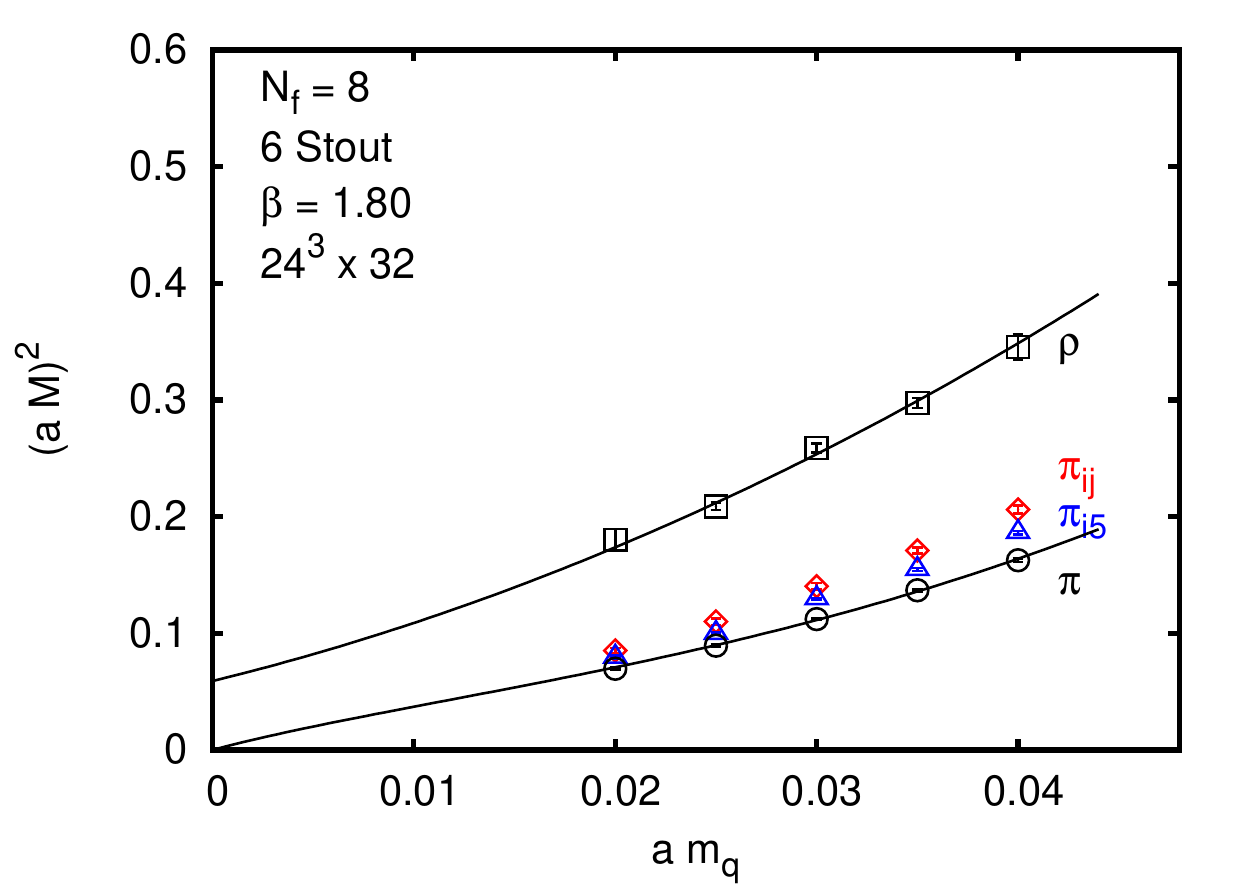}&
\includegraphics[height=5.3cm]{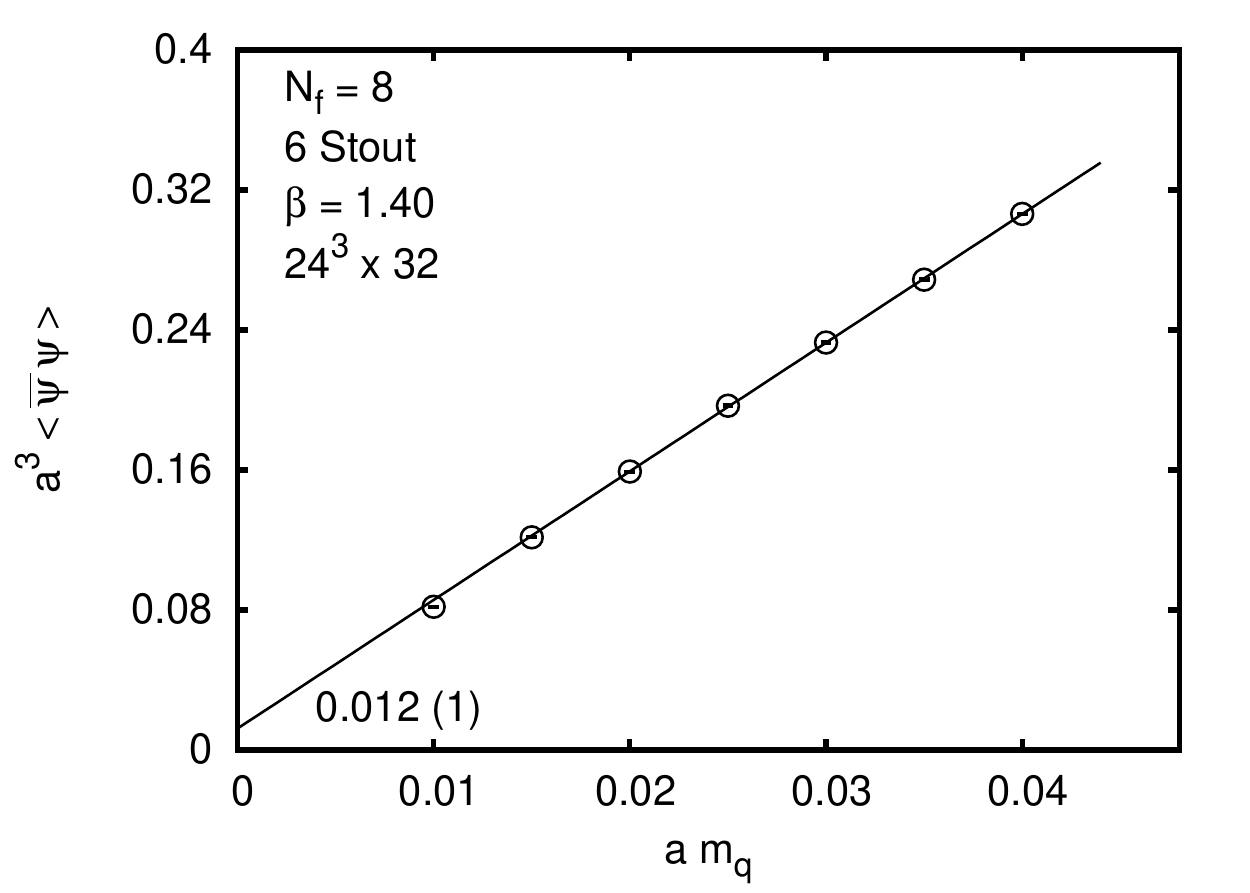}
\end{tabular}
\caption{The Goldstone spectrum and chiral fits are shown for $N_f=8$
simulations with lattice size $24^3\times 32$. The left column shows the
pseudo-Goldstone spectrum with decreasing taste breaking as the gauge coupling is
varied from $\beta=1.0$ to $\beta=1.8$.  The middle value at $\beta=1.4$ was
chosen in the top right figure with fitting range $a\cdot m_q=0.015-0.03$ of
the NLO chiral fit to $M^2_\pi/m_q$ which approaches $2B$ in the chiral limit.
The  middle figure on the right shows the $F_\pi$ data with no NLO fit far away
from the chiral limit. The bottom right figure is the linear fit to the chiral
condensate with fitting range $a\cdot m_q=0.02-0.04$. The physical fit
parameters $B,F,\Lambda_3$ are discussed in the text.}
\end{center}
\label{fig:Nf8}
\vskip -0.2in
\end{figure}

As we move to the $N_f=8$ p-regime simulations, we can clearly identify the
p-regime of the chirally broken phase as summarized in Fig.~\ref{fig:Nf8}. The
same lattice action and algorithm was used for the $N_f=8$ p-regime simulations
as introduced earlier for $N_f=4$. We can clearly identify the pseudo-Goldstone
spectrum which is separated from the technicolor scale of the $\rho$-meson.
Moving towards the continuum limit we observe at $\beta=1.4$ the split pion
spectrum collapsing toward the true Goldstone pion with a new distinct feature.
The true Goldstone pion, whose mass will vanish in the $a\cdot m_q=0$ limit at
fixed lattice spacing, and two additional split pseudo-Goldstone states appear with
considerably different slopes in Fig.~\ref{fig:Nf8} as $m_q$ increases. For
small $a \cdot m_q$ we find the pseudo-Goldstone spectrum collapsed at fixed gauge
coupling. Apparently the NLO operators, the last two terms in
Eq.~(\ref{eq:pimassform}), have a stronger effect on the spectra relative to
leading-order taste breaking operators, the generalization of those from
$N_f=4$ to $N_f=8$ as discussed in Section 2. This somewhat unexpected and
unexplained trend is observed for $N_f > 8$ as well.

We analyzed the $\chi{\rm SB}$ pattern within staggered perturbation theory in
its generalized form beyond four flavors~\cite{Aubin:2003mg,Aubin:2003uc}. The
simultaneous chiral fit of $M^2_\pi/m_q$ and $F_\pi$ based on
Eqs.~(\ref{eq:Mpi}-\ref{eq:FpiL}) cannot be done at $N_f=8$ within the reach of
the largest lattice sizes we deploy since the value of $F \cdot L_s$ is too
small even at $L_s=24$, for gauge couplings where taste breaking drops
to an acceptable level. The chiral fit of $B,F,\Lambda_3$ to $M^2_\pi/m_q$,
shown at the top right of Fig.~\ref{fig:Nf8}, is based on Eq.~(\ref{eq:Mpi})
only since the $F_\pi$ data points are outside the convergence range of the
chiral expansion. Much larger lattices are required to drop down in $m_q$ to
the region where the simultaneous fit could be made, while maintaining some
control over finite volume and taste breaking effects. The finite volume
corrections were negligible in the fitted $a\cdot m_q$ range and
Eqs.~(\ref{eq:MpiL},\ref{eq:FpiL}) were not needed. 

At $\beta=1.4$ the fitted value of $B$ is $a\cdot B=2.6(3)$ in lattice units with
$a\cdot F=0.0166(9)$ and $a\cdot \Lambda_3=0.48(5)$ also fitted. The fitted
$\rho$-mass in the chiral limit is $a\cdot M_\rho=0.27(2)$ with
$M_\rho/F=17(1)$. The fitted value of $B/F = 158(17)$ is not very reliable
but indicates significant enhancement of the chiral condensate from its $N_f=4$
value without including renormalization scale effects. The chiral condensate
$\langle \bar{\psi}\psi\rangle$ summed over all flavors is dominated by the
linear term in $m_q$ from UV contributions. The linear fit gives $\langle
\bar{\psi}\psi\rangle=0.012(1)$ in the chiral limit which differs from the GMOR
relation of $\langle \bar{\psi}\psi\rangle=8F^2B$ by about a factor of two with
$8F^2B=0.0058(8)$ fitted. There are several sources of this disagreement which
were addressed for the $N_f=4$ case earlier. The chiral log in $\langle
\bar{\psi}\psi\rangle$ will bring further down the true fitted value in the
chiral limit. Our volumes are not large enough yet to attempt a sensible chiral
log fit to the condensate at small $a\cdot m_q$ values. Finite volume squeezing
effects distort the consistency of the results in our limited range of
simulation volumes. Similar observations should also be noted when the RMT
analysis is applied in the $\eps$-regime.

\section {Goldstone spectrum and $\chi{\rm SB}$ from simulations at $\bf  N_f=9 $  in the p-regime}

\begin{figure}[ht!]
\begin{center}
\begin{tabular}{cc}
\includegraphics[height=5.3cm]{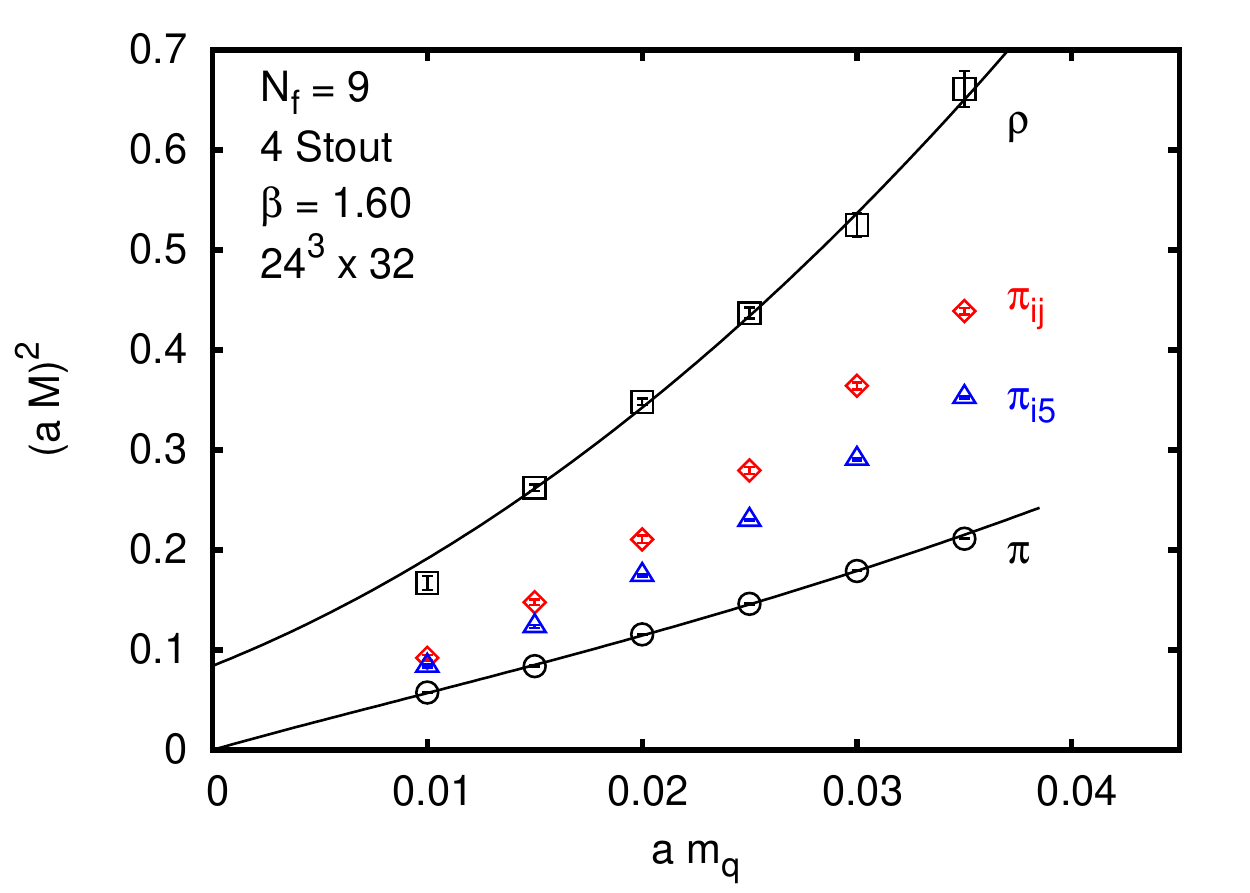}&
\includegraphics[height=5.3cm]{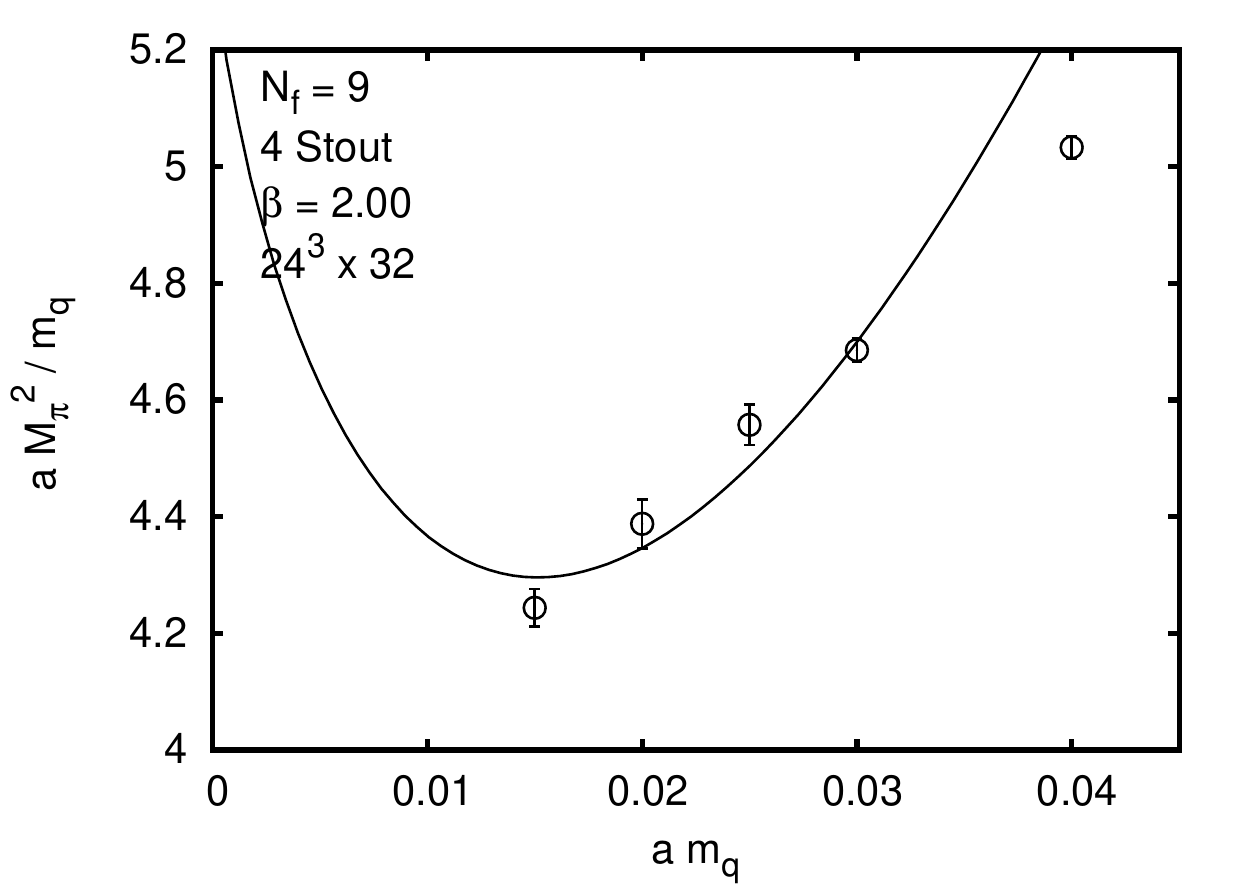}\\
\includegraphics[height=5.3cm]{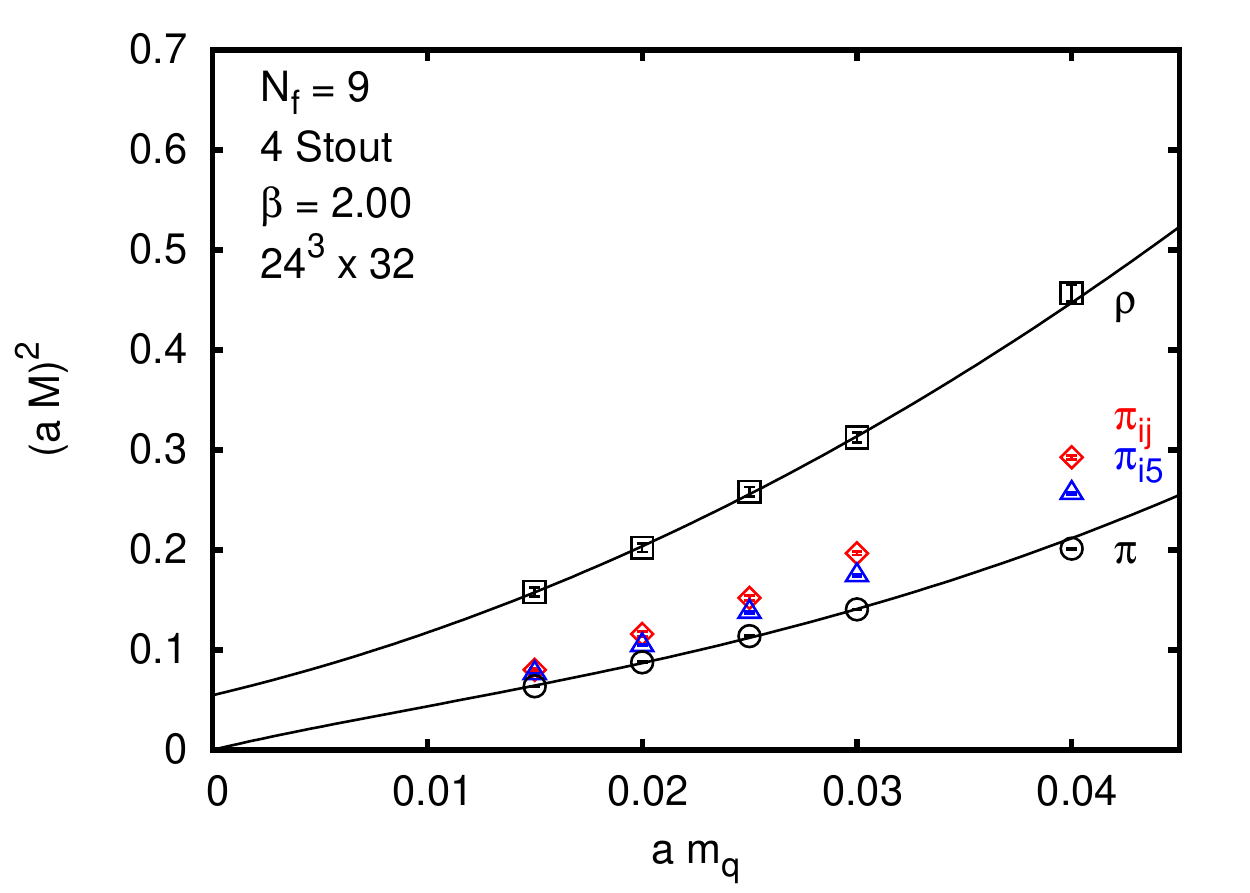}&
\includegraphics[height=5.3cm]{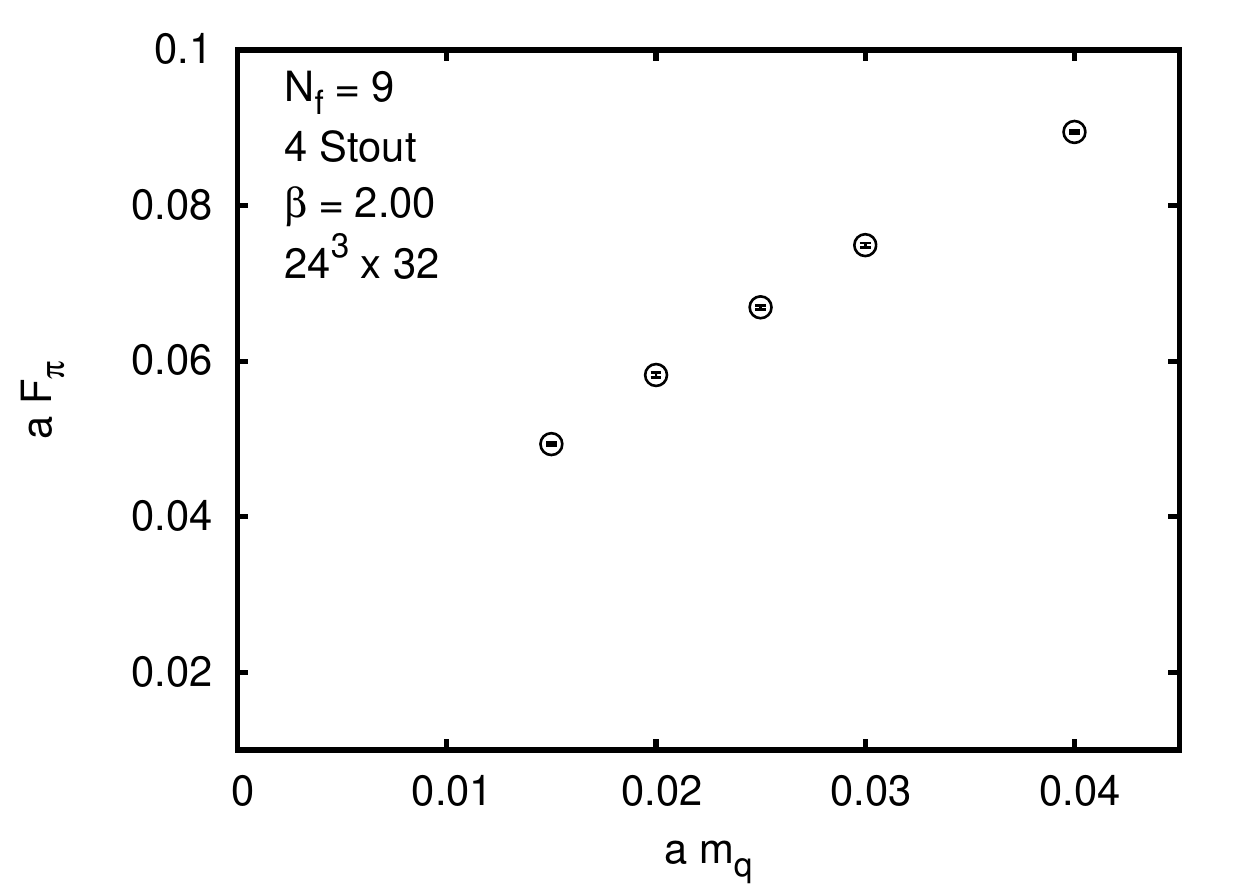}\\
\includegraphics[height=5.3cm]{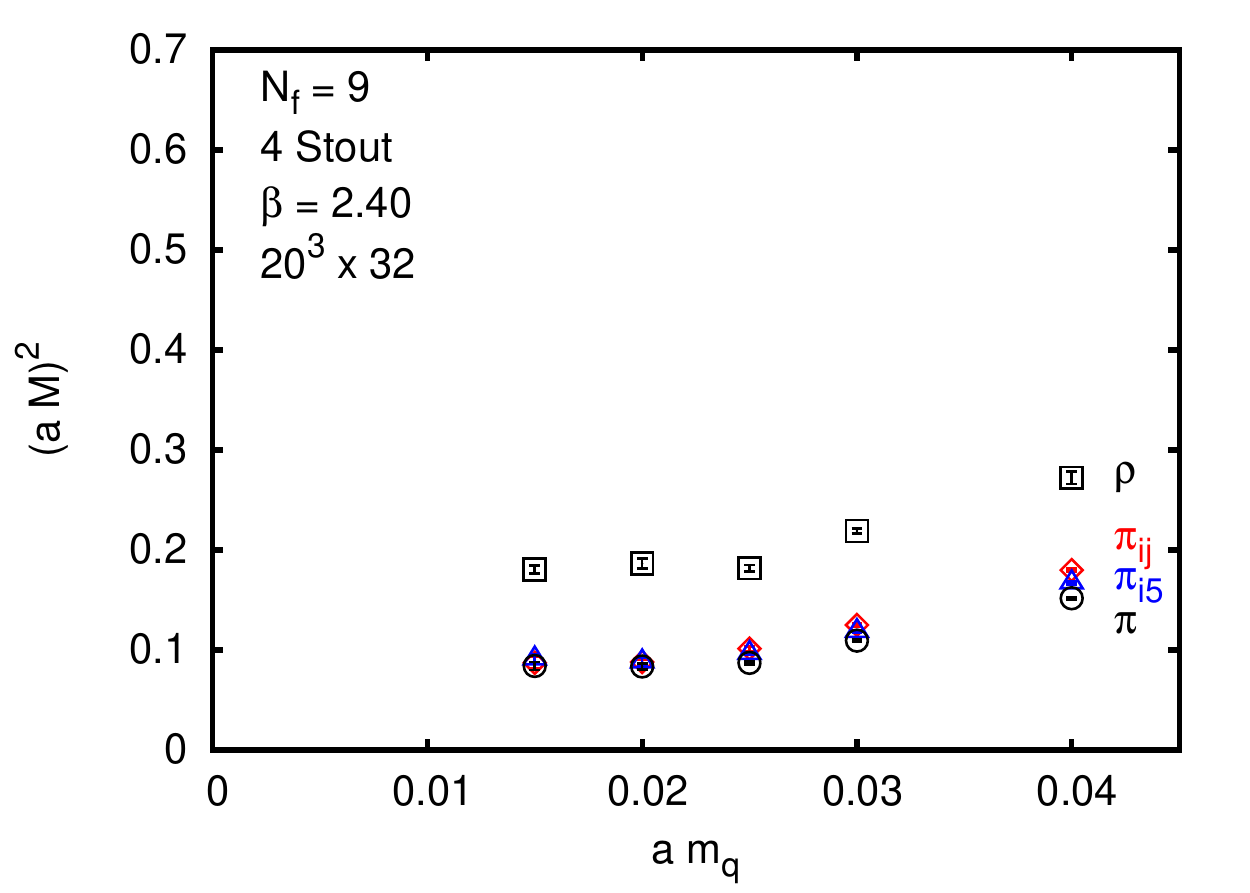}&
\includegraphics[height=5.3cm]{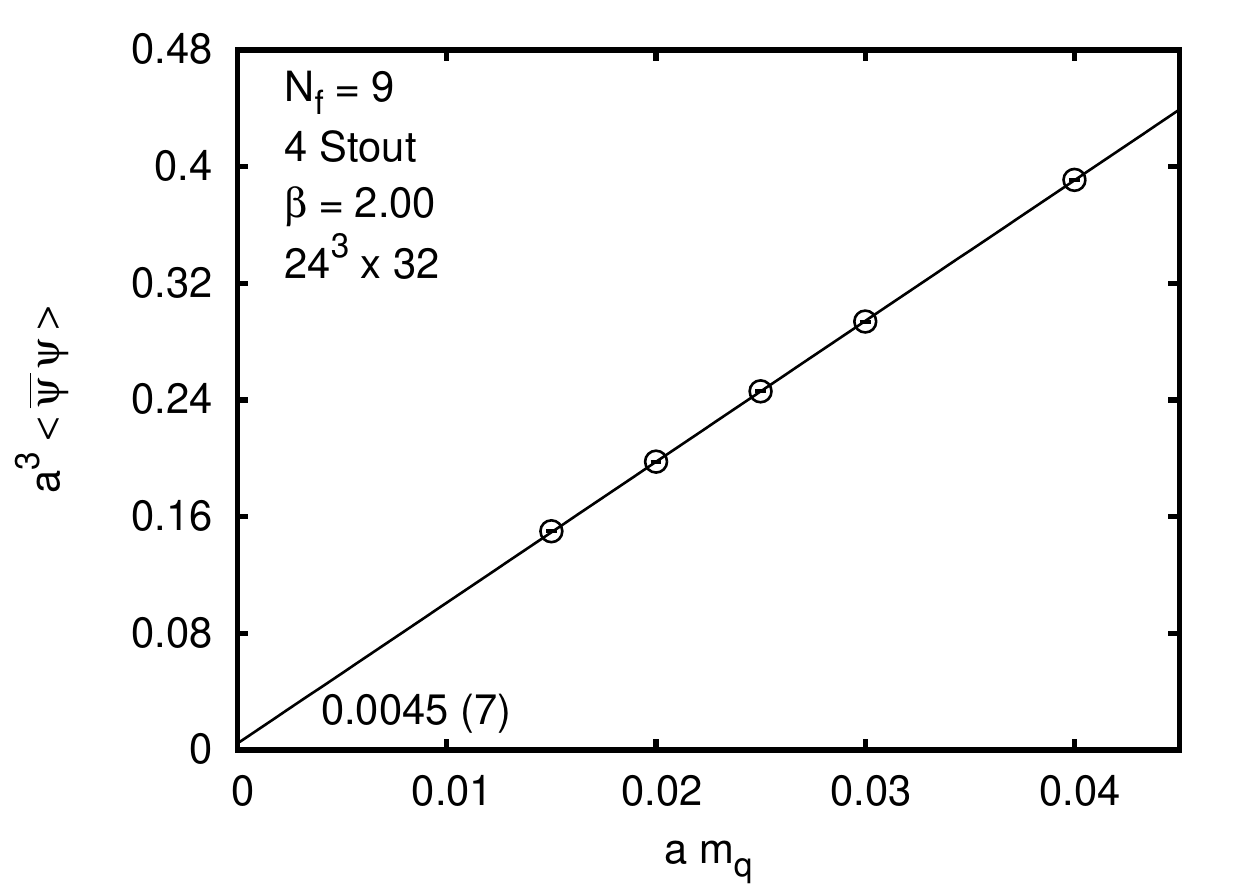}
\end{tabular}
\caption{The pseudo-Goldstone spectrum and chiral fits are shown for $N_f=9$
simulations with lattice size $24^3\times 32$. The left column shows the
pseudo-Goldstone spectrum with decreasing taste breaking as the gauge coupling is
varied from $\beta=1.6$ to $\beta=2.4$. Although the bottom figure on the left
at $\beta=2.4$ illustrates the continued restoration of taste symmetry, the
volume is too small for the Goldstone spectrum.  The middle value at
$\beta=2.0$ was chosen in the top right figure with fitting range $a\cdot
m_q=0.015-0.03$ of the NLO chiral fit to $M^2_\pi/m_q$ which approaches $2B$ in
the chiral limit.  The  middle figure on the right shows the $F_\pi$ data with
no NLO fit far away from the chiral limit. The bottom right figure is the
linear fit to the chiral condensate with fitting range $a\cdot m_q=0.02-0.04$.
The physical fit parameters $B,F,\Lambda_3$ are discussed in the text. Four
stout steps were used in all $N_f=9$ simulations.}   
\end{center}
\label{fig:Nf9}
\vskip -0.2in
\end{figure}

We had two motivations for the $N_f=9$ simulation reported here. We wanted to
see whether the rooting procedure (being applied in our project with two
fermions in the sextet representation) will present some unexpected changes in
the analysis and we were also looking for the continued trends in the $\chi{\rm
SB}$ pattern. We could not find any noticeable effect from the rooting
procedure and the symmetry breaking pattern was consistent with the $N_f=8$
simulations.

As shown in Fig.~\ref{fig:Nf9} the Goldstone spectrum is still clearly separated
from the technicolor scale of the $\rho$-meson. The true Goldstone pion and two
additional split pseudo-Goldstone states are shown again with different slopes as $a\cdot
m_q$ increases. The trends and the underlying explanation are similar to the
$N_f=8$ case. The chiral fit to $M^2_\pi/m_q$ is shown based on
Eq.~(\ref{eq:Mpi}) only since the $F_\pi$ data points are outside the
convergence range of the chiral expansion. At $\beta=2.0$ the fitted value of $B$
is $a\cdot B=2.8(4)$ in lattice units with $a\cdot F=0.017(2)$ and $a\cdot
\Lambda_3=0.48(9)$ also fitted. The fitted $\rho$-mass in the chiral limit is
$a\cdot M_\rho=0.233(3)$ with $M_\rho/F=14(1)$. The fitted value of $B/F =
166(32)$ is not very reliable but comparable to the enhancement of the chiral
condensate found at $N_f=8$ without including renormalization scale effects.
Again, at fixed lattice spacing, the small chiral condensate $\langle
\bar{\psi}\psi\rangle$ summed over all flavors is dominated by the linear term
in $m_q$ from UV contributions. The linear fit gives $\langle
\bar{\psi}\psi\rangle=0.0045(7)$ in the chiral limit which differs from the
GMOR relation of $\langle \bar{\psi}\psi\rangle=9F^2B$ by about a factor of two
with $9F^2B=0.007(2)$ fitted. Open issues in the systematics are similar to
the $N_f=8$ case.

\section {Goldstone spectrum and $\chi {\rm SB}$ from simulations at $\bf  N_f=12$  in the p-regime}

\begin{figure}[ht!]
\begin{center}
\begin{tabular}{cc}
\includegraphics[height=5.3cm]{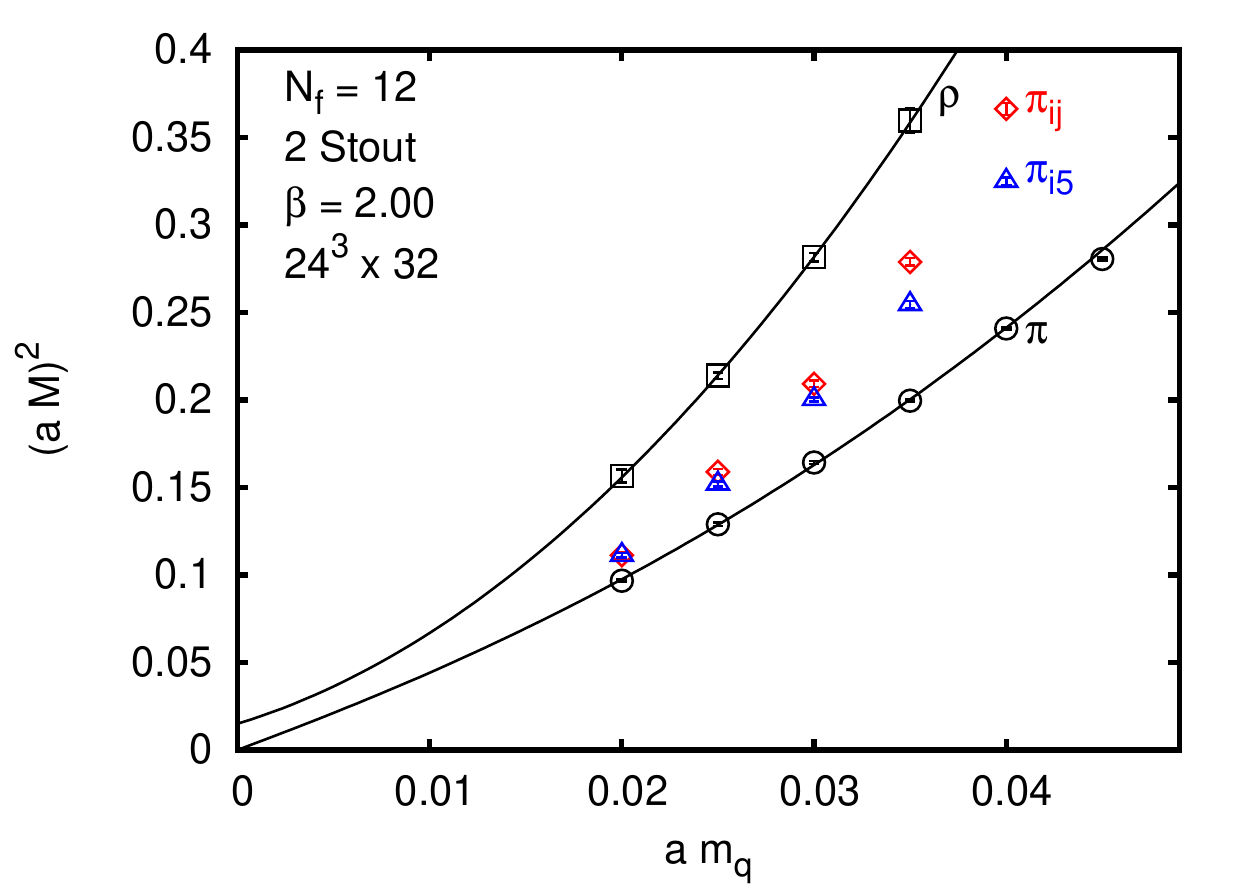}&
\includegraphics[height=5.3cm]{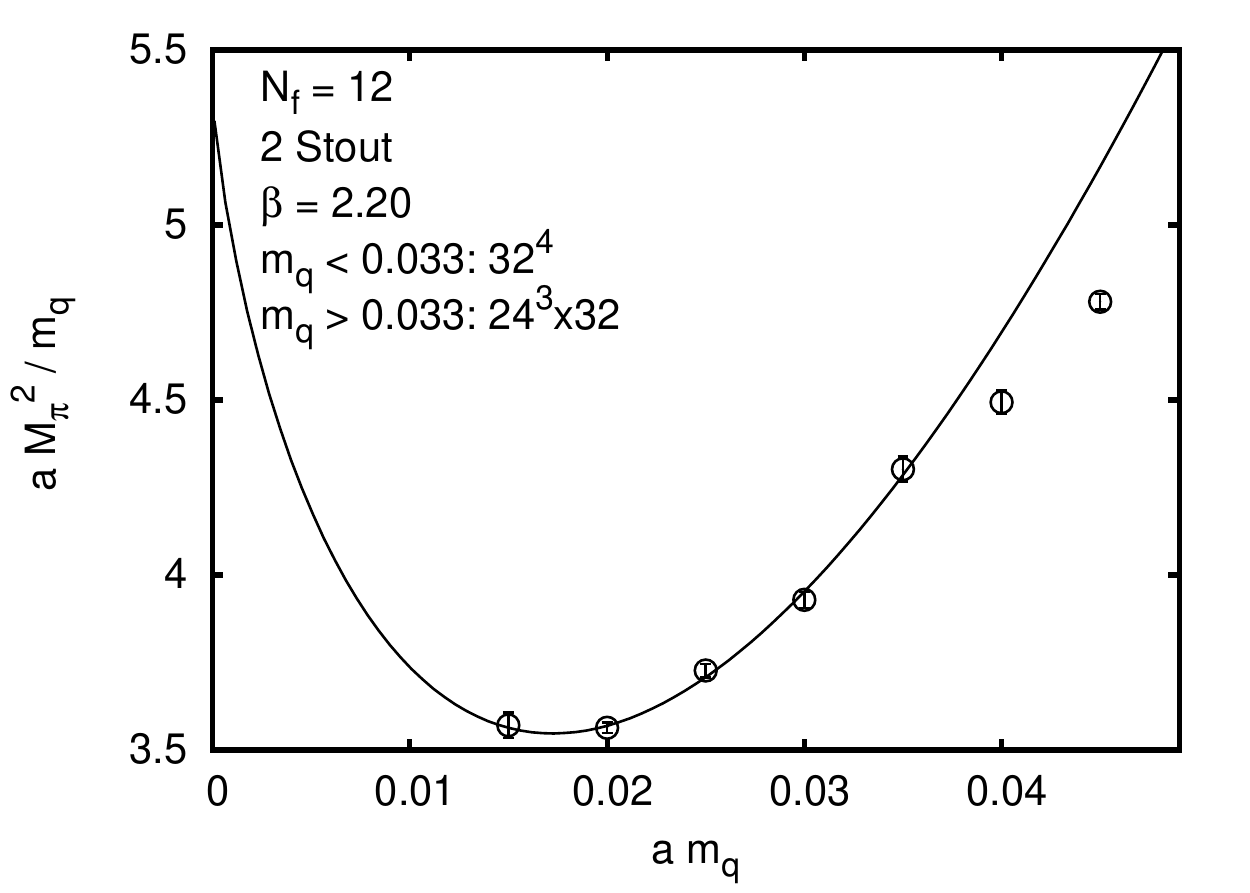}\\
\includegraphics[height=5.3cm]{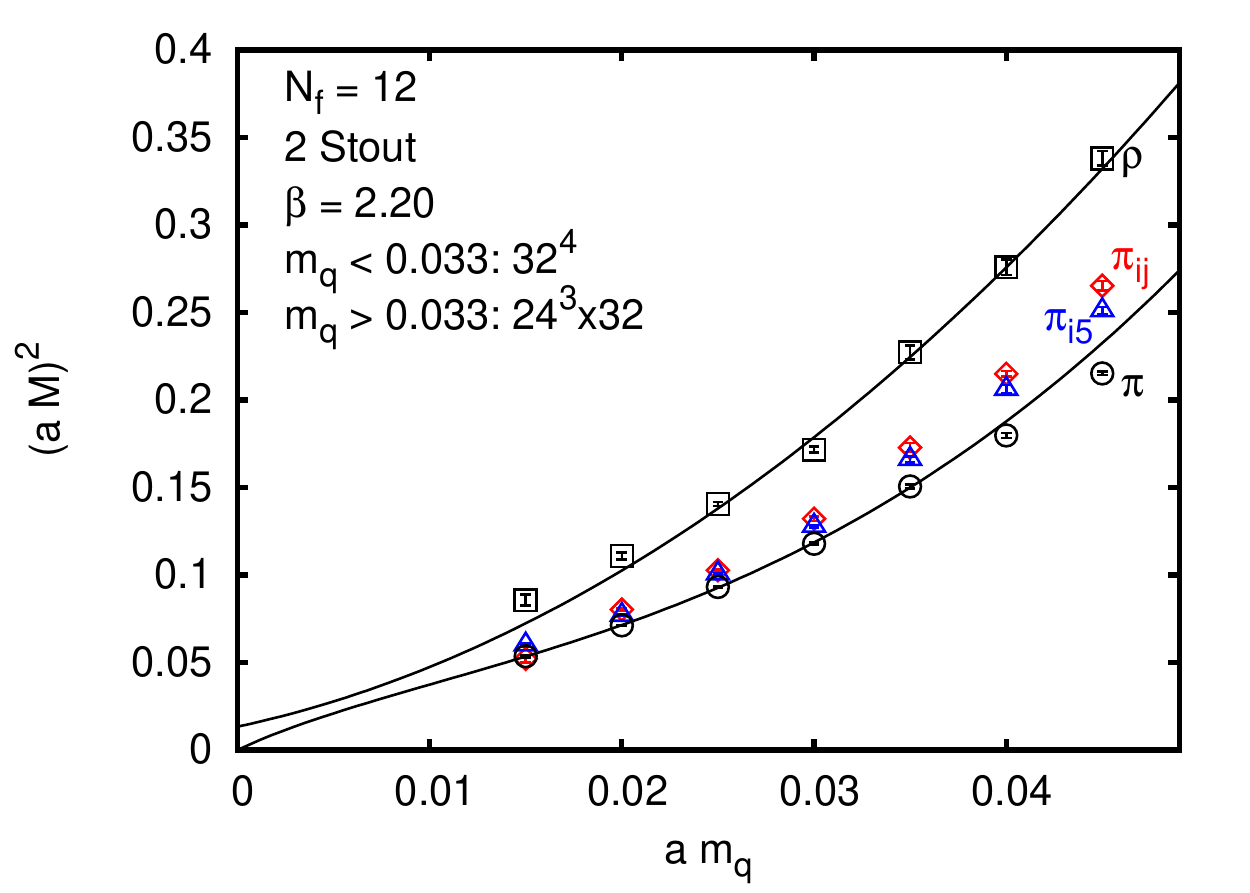}&
\includegraphics[height=5.3cm]{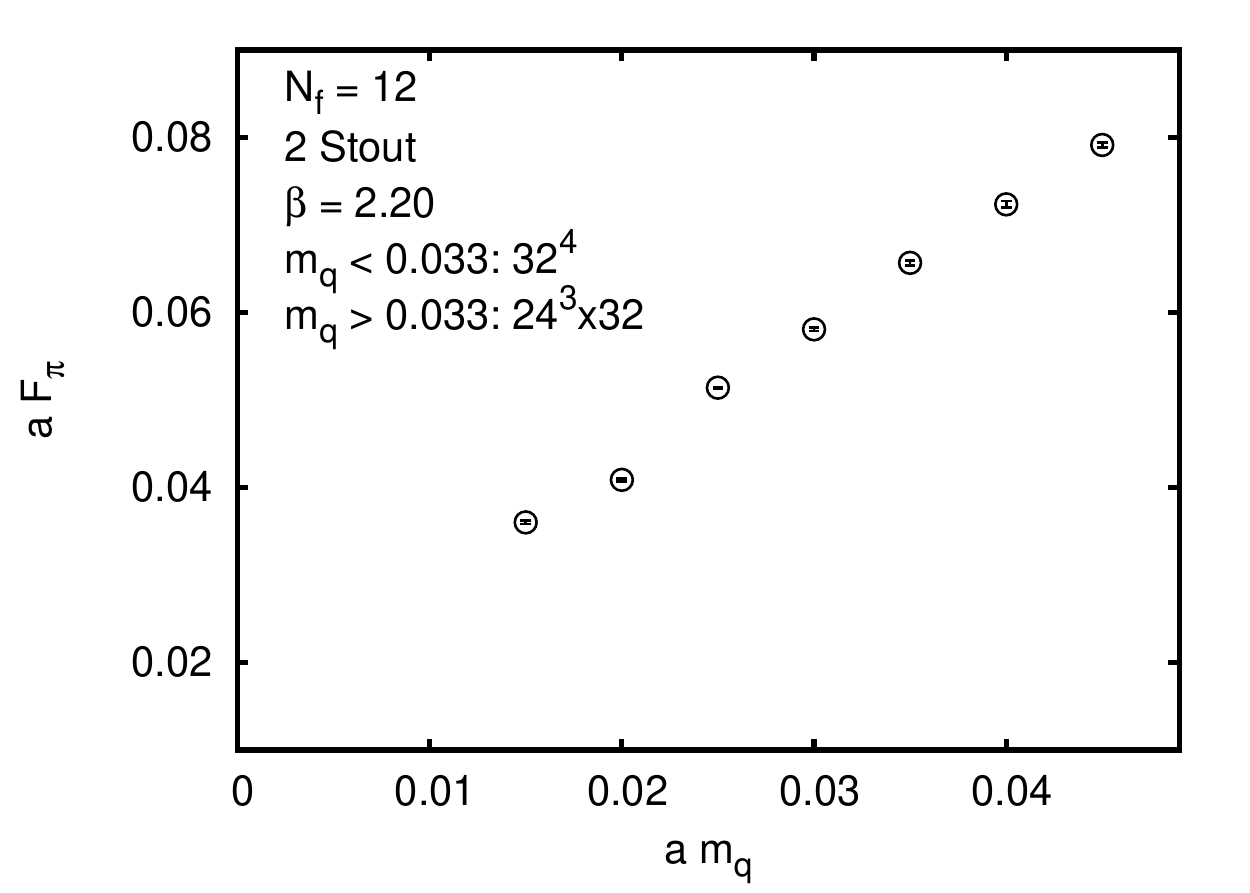}\\
\includegraphics[height=5.3cm]{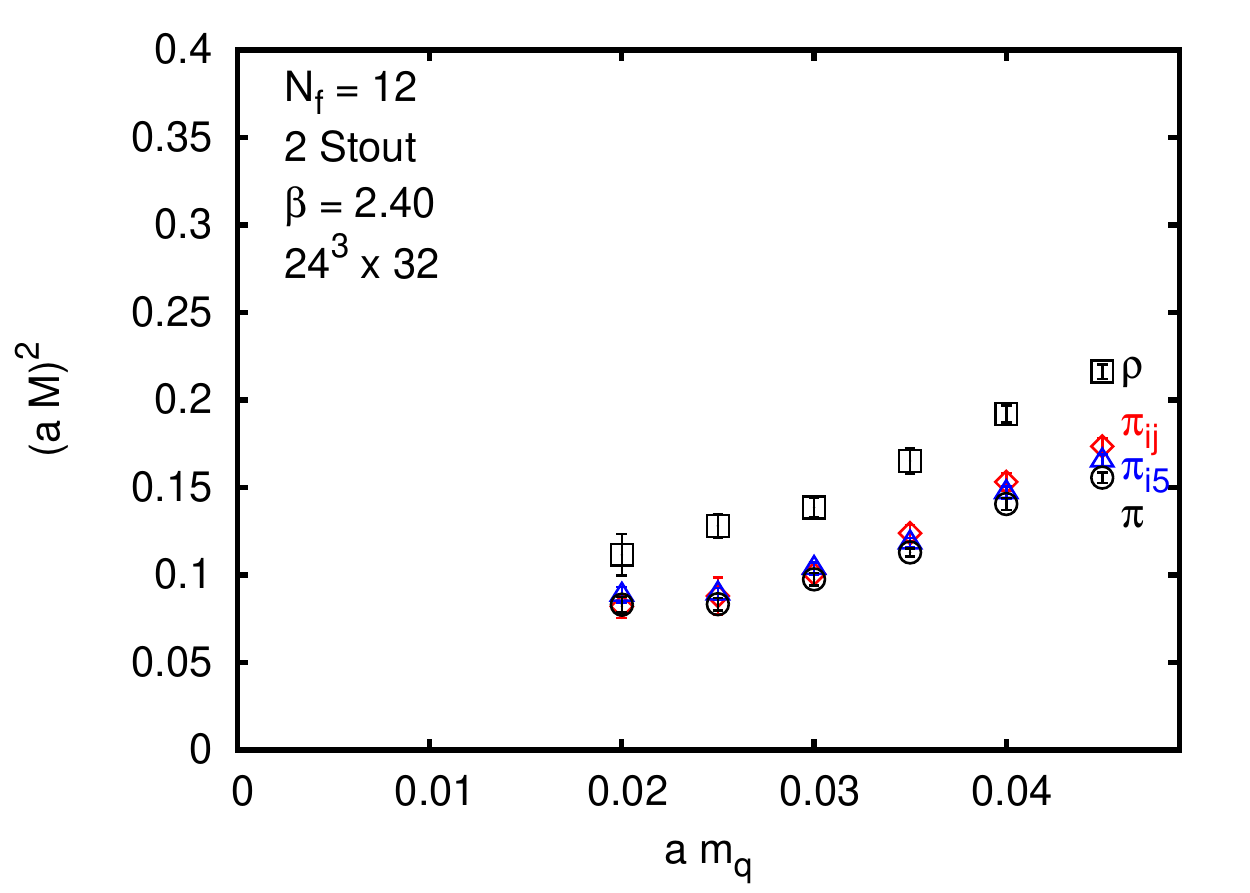}&
\includegraphics[height=5.3cm]{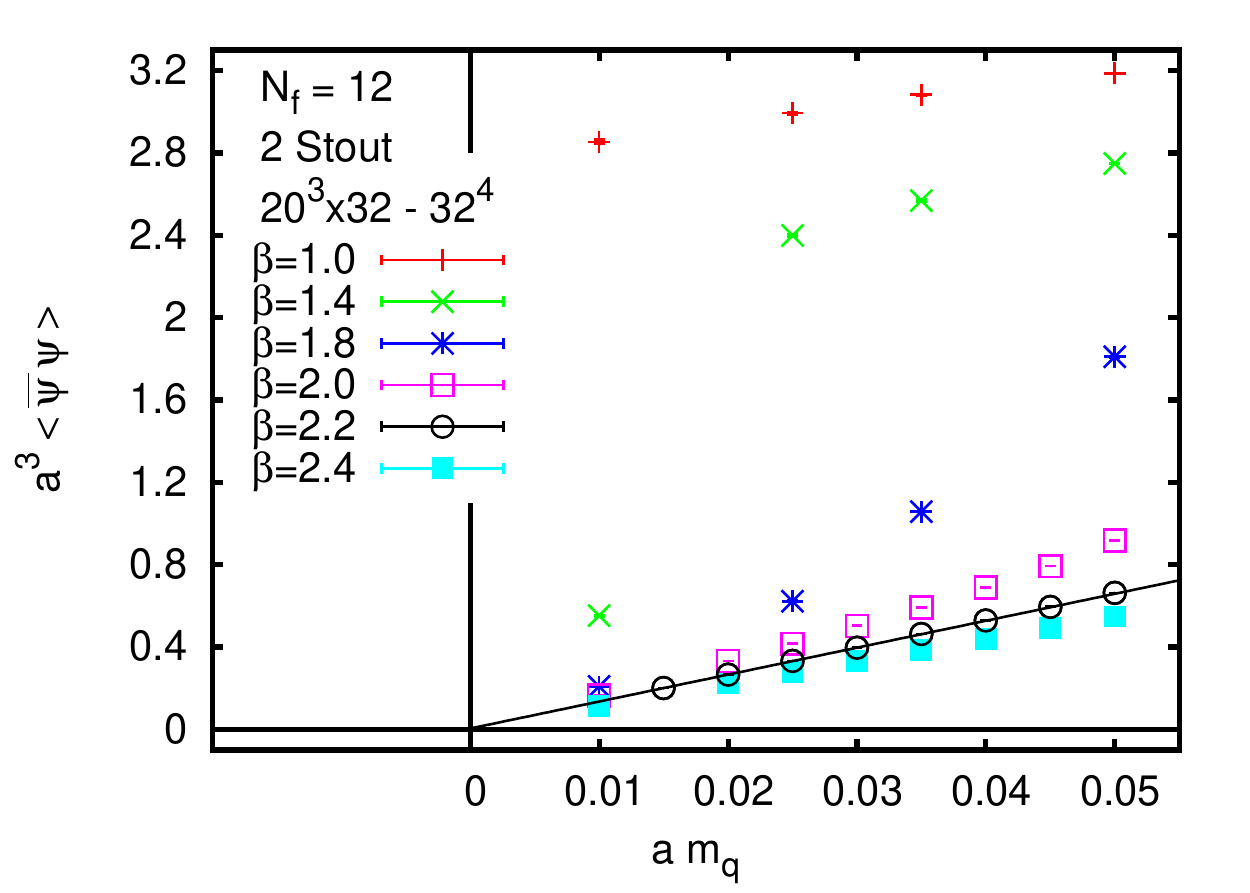}
\end{tabular}
\caption{The pseudo-Goldstone spectrum and chiral fits are shown for $N_f=12$
simulations with lattice size $24^3\times 32$ and $32^4$. The left column shows
the pseudo-Goldstone spectrum with decreasing taste breaking as the gauge coupling is
varied from $\beta=2.0$ to $\beta=2.4$. Although the bottom figure on the left
at $\beta=2.4$ illustrates the continued restoration of taste symmetry, the
volume is too small for the Goldstone spectrum.  The middle value at
$\beta=2.2$ was chosen in the top right figure with fitting range $a\cdot
m_q=0.015-0.035$ of the NLO chiral fit to $M^2_\pi/m_q$ which approaches $2B$
in the chiral limit.  The  middle figure on the right shows the $F_\pi$ data
with no NLO fit far away from the chiral limit. The bottom right figure, 
with its additional features discussed in the text, is the
linear fit to the chiral condensate with fitting range $a\cdot m_q=0.02-0.04$.
The physical fit parameters $B,F,\Lambda_3$ are discussed in the text. Two
stout steps were used in all $N_f=12$ simulations.}   
\end{center}
\label{fig:Nf12}
\vskip -0.2in
\end{figure}

Finally we move to the controversial $N_f=12$ case. We find here a similar
chiral symmetry breaking pattern as we found in the $N_f=8,9$ cases with
increased concerns about all the caveats presented before.  The Goldstone
spectrum remains  separated from the technicolor scale of the $\rho$-meson. The
true Goldstone pion and two additional split pseudo-Goldstone states are shown
again in Fig.~\ref{fig:Nf12} with different slopes as $a\cdot m_q$ increases.
The trends and the underlying explanation are  similar to the $N_f=8,9$ cases.
The chiral fit to $M^2_\pi/m_q$ shown at the top right side of
Fig.~\ref{fig:Nf12} is based on Eq.~(\ref{eq:Mpi}) only since the $F_\pi$ data
points are outside the convergence range of the chiral expansion.  At
$\beta=2.2$ the fitted value of  $B$ is $a\cdot B=2.7(2)$ in lattice units with
$a\cdot F=0.0120(1)$ and $a\cdot \Lambda_3=0.50(3)$ also fitted. The fitted
$\rho$-mass in the chiral limit is $a\cdot M_\rho=0.115(15)$ from $a\cdot
m_q=0.025-0.045$ with $M_\rho/F=10(1)$.  The fitted value of $B/F = 223(17)$ is
not very reliable but consistent with the  enhancement of the chiral condensate
found at $N_f=8,9$  without including renormalization scale effects.  Again, at
fixed lattice spacing, the small  chiral condensate $\langle
\bar{\psi}\psi\rangle$ summed over all flavors is dominated by the linear term
in $m_q$ from UV contributions.  The linear fit gives $\langle
\bar{\psi}\psi\rangle=0.0033(13)$ in the chiral limit which came out
unexpectedly close the GMOR relation of  $\langle \bar{\psi}\psi\rangle=12F^2B$
with $12F^2B=0.0046(4)$ fitted. Issues and concerns in the systematics are
similar to the $N_f=8,9$ cases.

In summary, we have shown strong evidence that  according to p-regime tests the
$N_f=4,8,9,12$ systems all exhibit in the Goldstone and hadron spectra broken
chiral symmetry close to the continuum limit.  There are some important
features of  the $N_f=12$ analysis which suggest that the model is not only in
the $\chi{\rm SB}$ phase but also close to slow walking of the renormalized
gauge coupling. The bottom right of Fig.~\ref{fig:Nf12} shows the crossover in
the chiral condensate from strong coupling to the weak coupling regime in the
relevant range of $m_q$. In combination with the nearly degenerate Goldstone
spectrum we find it quite suggestive that around  $\beta=2.2$ we are close to
continuum behavior.  In addition, we observe that the fitted value of the
$\rho$-mass in the chiral limit hardly changes in this region as the gauge
coupling is varied (at $\beta=2.0$ we fit $a\cdot M_\rho=0.123(10)$).  If
confirmed on larger lattices, this could be a first hint of a slowly changing
gauge coupling close to the conformal window.  Currently we are investigating
the important $N_f=12$ model on larger lattices to probe the possible influence
of unwanted squeezing effects on the spectra. This should also clarify the mass splitting
pattern of the $\rho$ and $A_1$ states we are seeing in the chiral limit as $N_f$ is varied.

Our findings at $N_f=12$ are in disagreement with~\cite{Appelquist:2007hu,Appelquist:2009ty}.
Lessons from the Dirac spectra and RMT to complement p-regime tests are
discussed in the next section.
\begin{figure*}[ht]
\begin{center}
\begin{tabular}{cc}
\includegraphics[height=5.1cm]{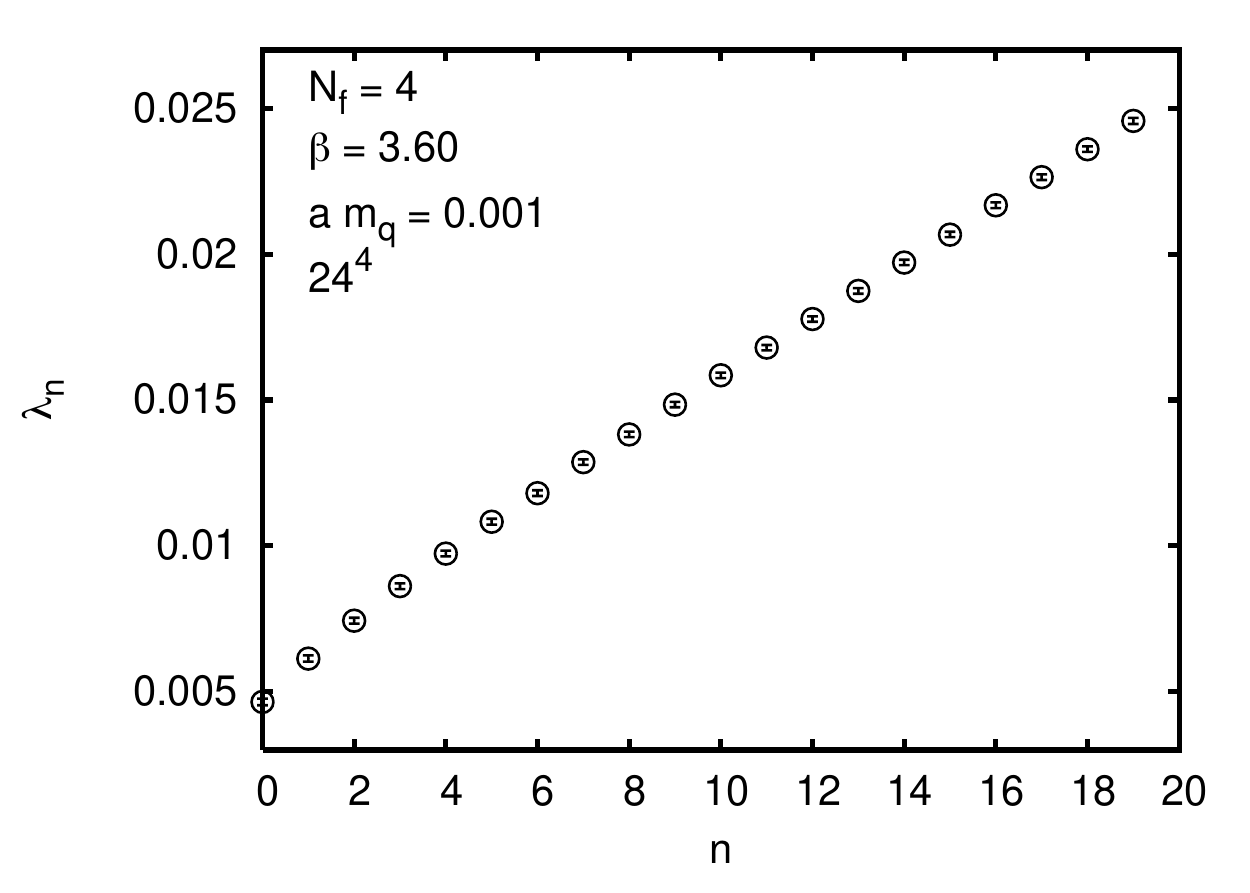}&
\includegraphics[height=5.1cm]{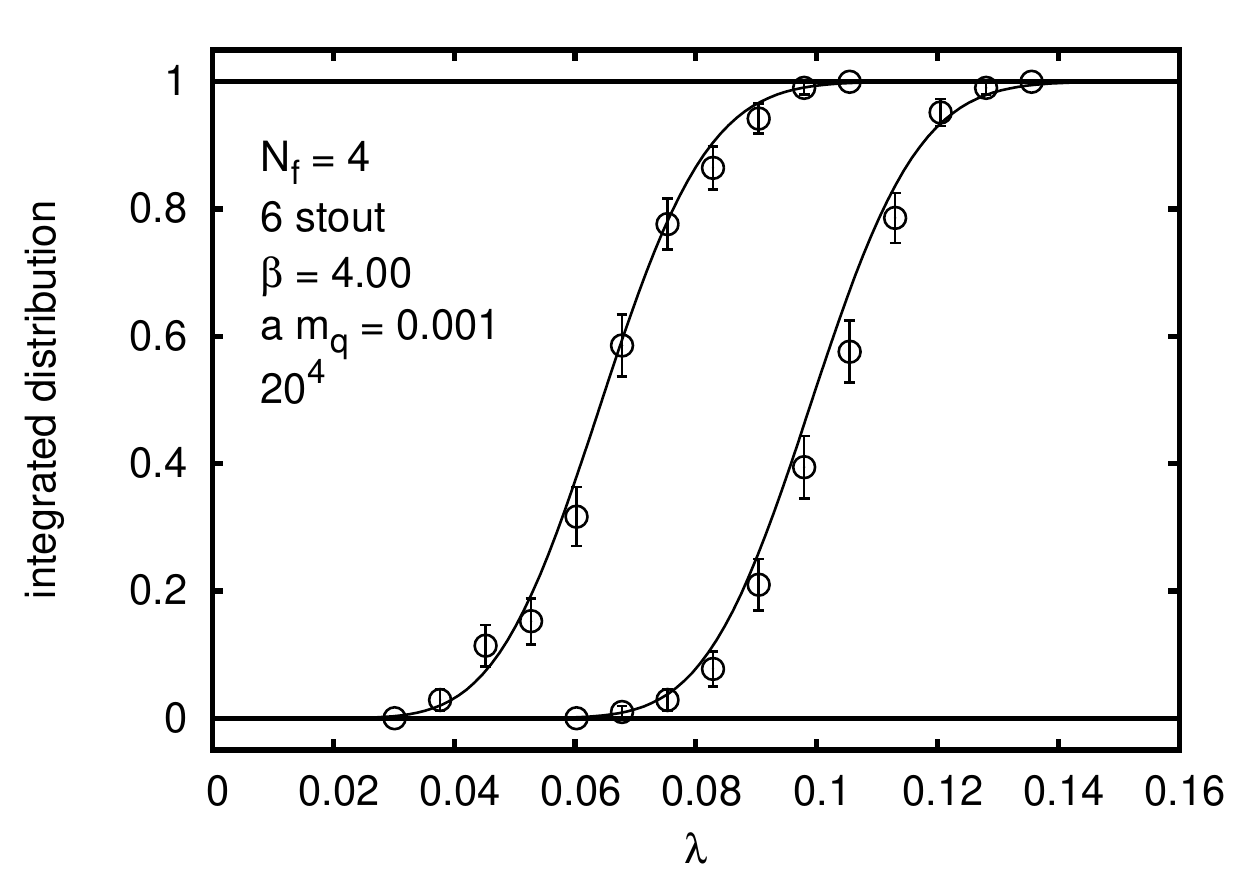}\\
\includegraphics[height=5.1cm]{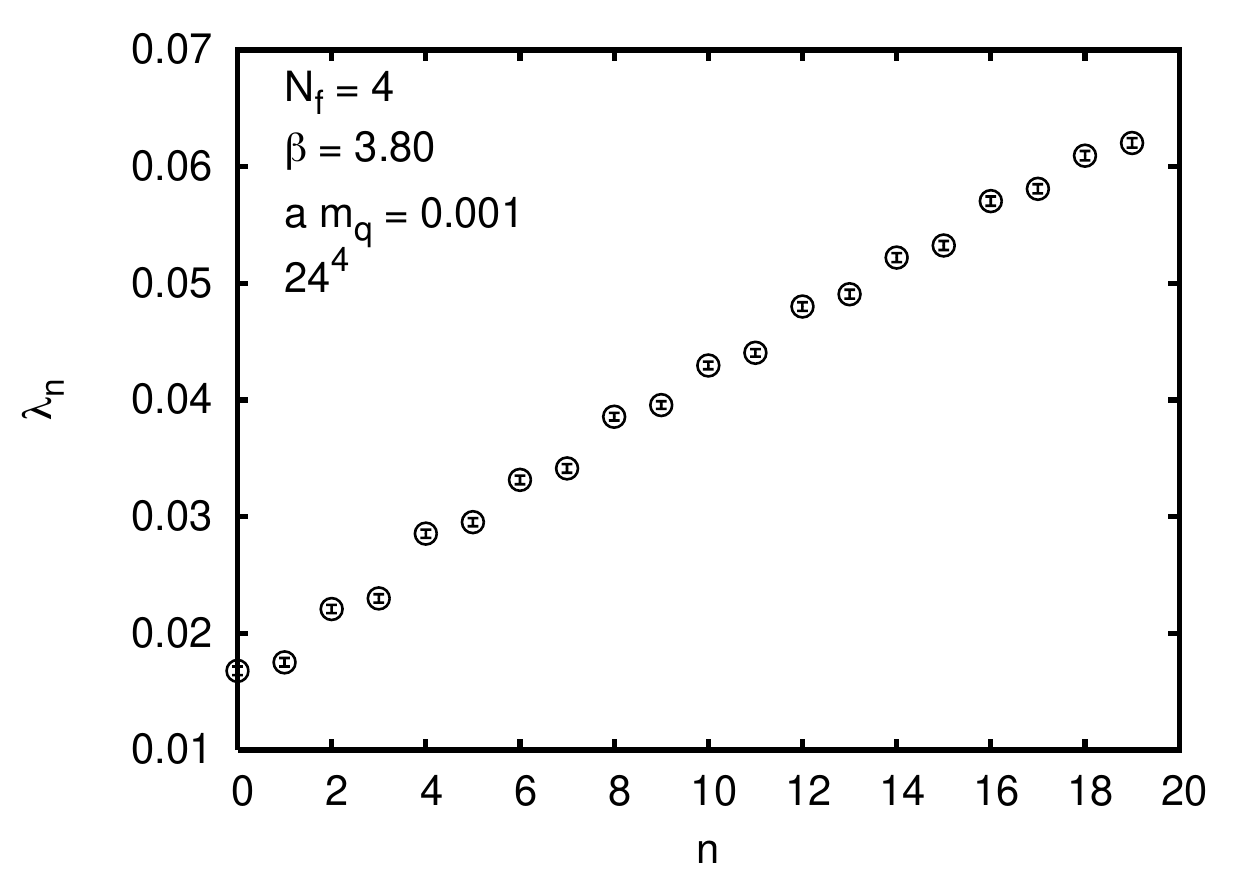}&
\includegraphics[height=5.1cm]{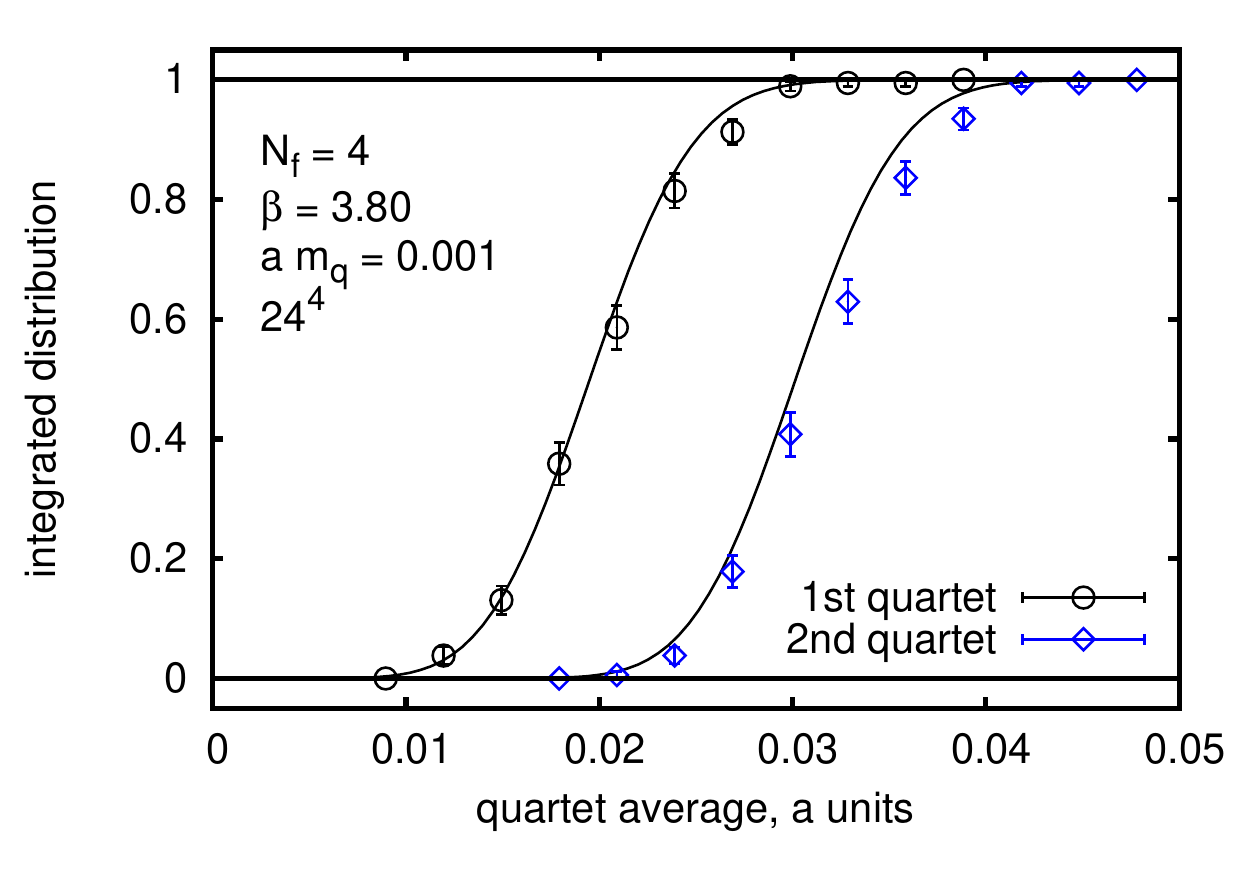}\\
\includegraphics[height=5.1cm]{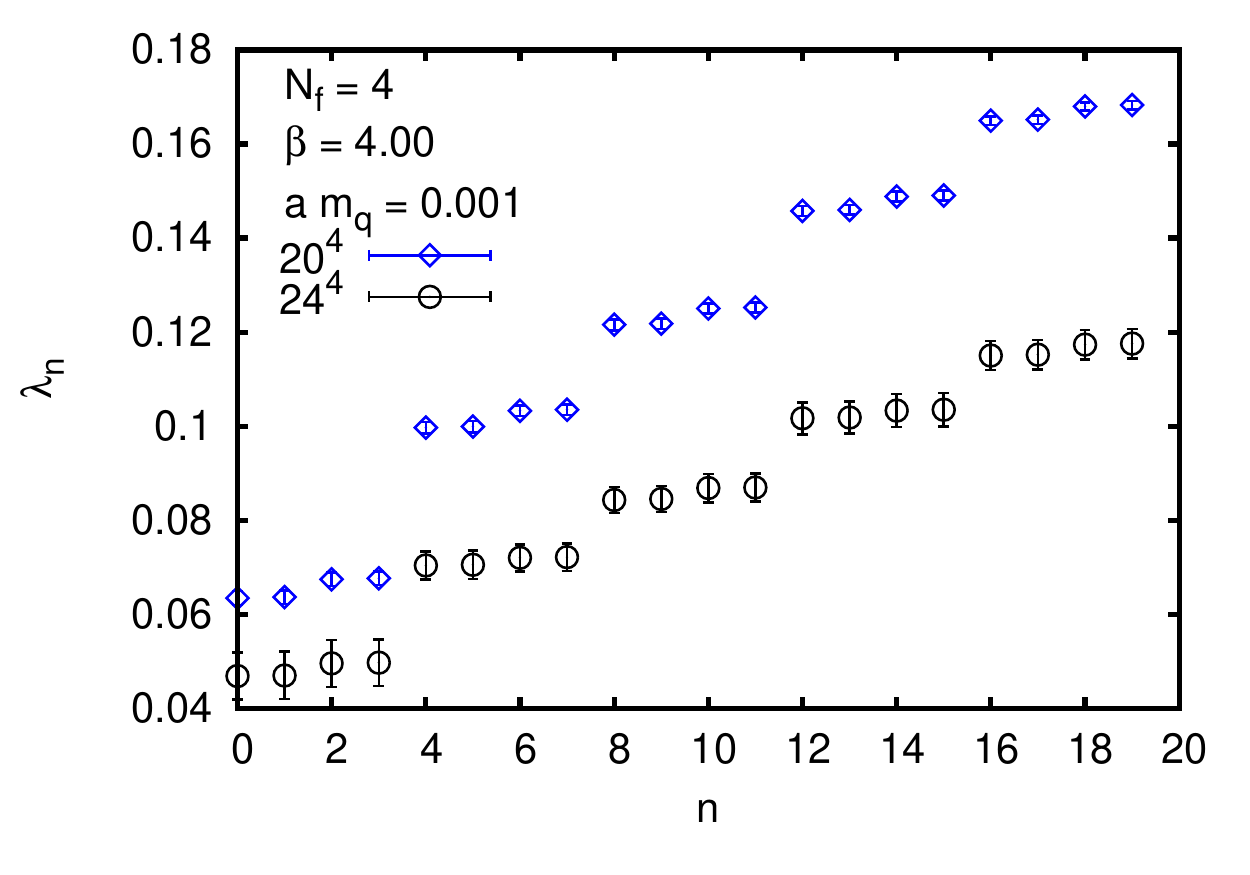}&
\includegraphics[height=5.1cm]{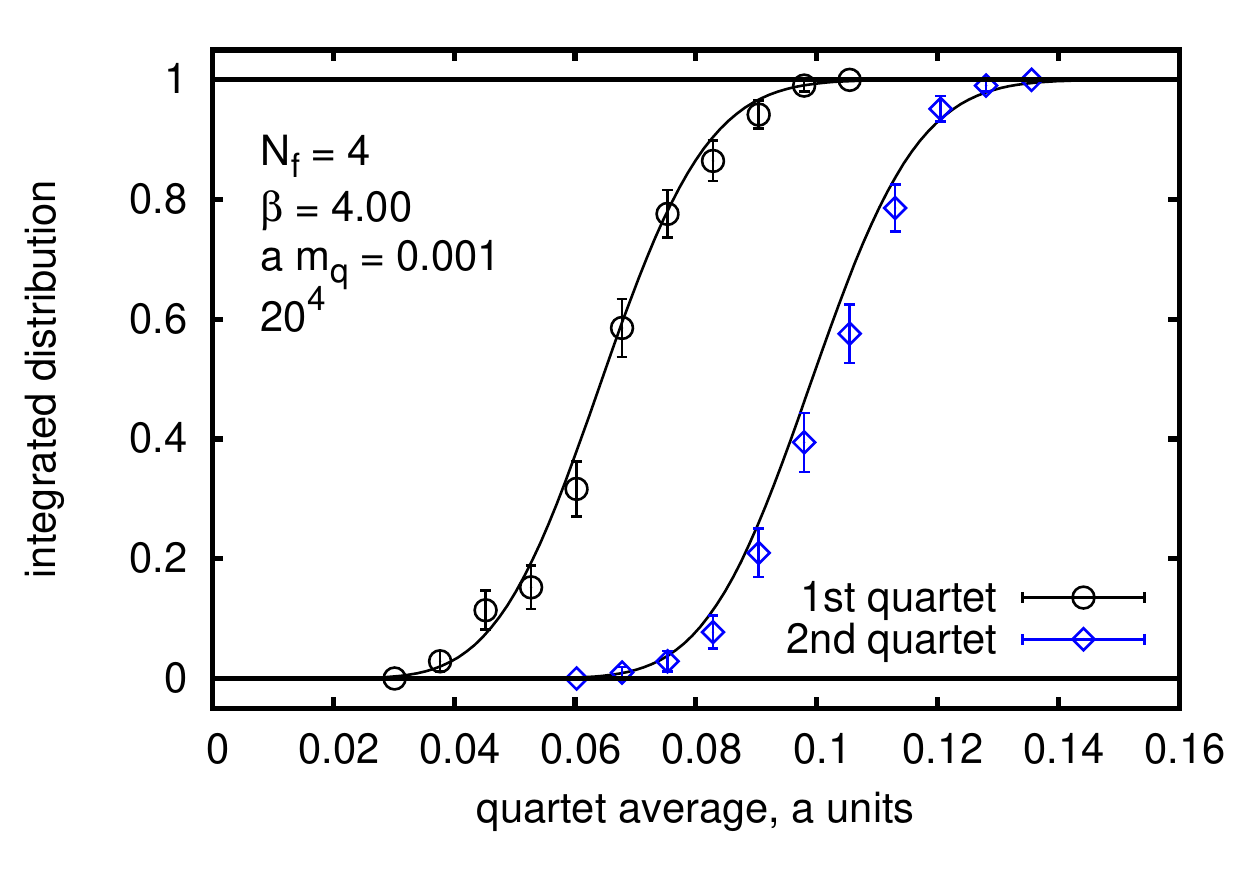}
\end{tabular}
\end{center}
\caption{From simulations at $N_f=4$ the first column shows the approach to
quartet degeneracy of the spectrum as $\beta$ increases.  The second column
shows the integrated distribution of the two lowest quartets averaged. The
solid line compares this procedure to RMT with $N_f=4$.}   
\label{fig:rmt}
\vskip -0.1in
\end{figure*}
\section{Epsilon regime, Dirac spectrum and RMT}

If the bare parameters of a gauge theory are tuned to the $\eps$-regime
in the chirally broken phase, the low-lying Dirac spectrum follows the
predictions of  random matrix theory.  The corresponding random matrix model is
only sensitive to the pattern of chiral symmetry breaking, the topological
charge and the rescaled fermion mass once the eigenvalues are also rescaled by
the same factor $\Sigma_{cond} V$.  This idea has been confirmed in various
settings both in quenched and fully dynamical simulations. The same method is
applied here to nearly conformal gauge models.

The connection between the eigenvalues $\lambda$ of the Dirac operator and
chiral symmetry breaking is  given in the Banks-Casher
relation~\cite{Banks:1979yr},
$$
\Sigma_{cond} = - \langle \bar{\Psi} \Psi \rangle = \lim_{\lambda \rightarrow
  0} \lim_{m \rightarrow 0} \lim_{V \rightarrow \infty} \frac{\pi
  \rho(\lambda)}{V},
$$
where $\Sigma_{cond}$ designates the quark condensate normalized to a single
flavor.  To generate a non-zero density $\rho(0)$, the smallest eigenvalues
must become densely packed as the volume increases, with an eigenvalue spacing
$\Delta \lambda \approx 1/\rho(0) = \pi/(\Sigma_{cond} V)$. This allows a crude
estimate of the quark condensate $\Sigma_{cond}$. One can do better by
exploring the $\eps$-regime: If chiral symmetry is spontaneously broken,
tune the volume and quark mass such that
$
\frac{1}{F_\pi} \ll L \ll \frac{1}{M_\pi},
$
so that the Goldstone pion is much lighter than the physical value, and finite volume
effects are dominant as we discussed in Section 2. The chiral Lagrangian of
Eq.~(\ref{eq:Lagrange}) is dominated by the zero-momentum mode from the mass
term and all kinetic terms are suppressed. In this limit, the distributions of
the lowest eigenvalues are identical to those of random matrix theory, a theory
of large matrices obeying certain
symmetries~\cite{Shuryak:1992pi,Damgaard:2003nm,Verbaarschot:2000dy}.  To
connect with RMT, the eigenvalues and quark mass are rescaled as $z = \lambda
\Sigma_{cond} V$ and $\mu = m _q\Sigma_{cond} V$, and the eigenvalue
distributions also depend on the topological charge $\nu$ and the number of
quark flavors $N_f$. RMT is a very useful tool to calculate analytically all of
the eigenvalue distributions~\cite{Damgaard:2000ah}. The eigenvalue
distributions in various topological sectors are measured via lattice
simulations, and via comparison with RMT, the value of the condensate
$\Sigma_{cond}$ can be extracted. 

After we generate large thermalized ensembles, we calculate the lowest twenty
eigenvalues of the Dirac operator using the PRIMME package~\cite{primme}. In
the continuum limit, the staggered eigenvalues form degenerate quartets, with
restored taste symmetry. The first column of  Fig.~\ref{fig:rmt} shows the change
in the eigenvalue structure for $N_f=4$ as the coupling constant is varied.  At
$\beta=3.6$ grouping into quartets is not seen, the Goldstone pions are somewhat still split,
and staggered perturbation theory is just beginning to kick in. At $\beta=3.8$
doublet pairing appears and at $\beta=4.0$ the quartets are nearly degenerate.
The Dirac spectrum is collapsed as required by the Banks-Casher relation.  In
the second column we show the integrated distributions of the two lowest
eigenvalue quartet averages,
\begin{equation}
\int_0^{\lambda} p_k(\lambda') d\lambda', \hspace{0.5cm} k=1,2
\end{equation} 
which is only justified close to quartet degeneracy. All low eigenvalues are
selected with zero topology.  To compare with RMT, we vary $\mu=m
_q\Sigma_{cond} V$ until we satisfy 
\begin{equation}
\frac{\langle \lambda_1 \rangle_{\rm sim}}{m} = \frac{\langle z_1
  \rangle_{\rm RMT}}{\mu},
\label{eq:rmt}
\end{equation}
where $\langle \lambda_1 \rangle_{\rm sim}$ is the lowest quartet average from
simulations and the RMT average $\langle z \rangle_{\rm RMT}$ depends
implicitly on $\mu$ and $N_f$. With this optimal value of $\mu$, we can predict
the shapes of $p_k(\lambda)$ and their integrated distributions, and compare to
the simulations. The agreement with the two lowest integrated RMT eigenvalue
shapes is excellent for the larger $\beta$ values.

\begin{figure}[h!]
\begin{center}
\includegraphics[height=7cm]{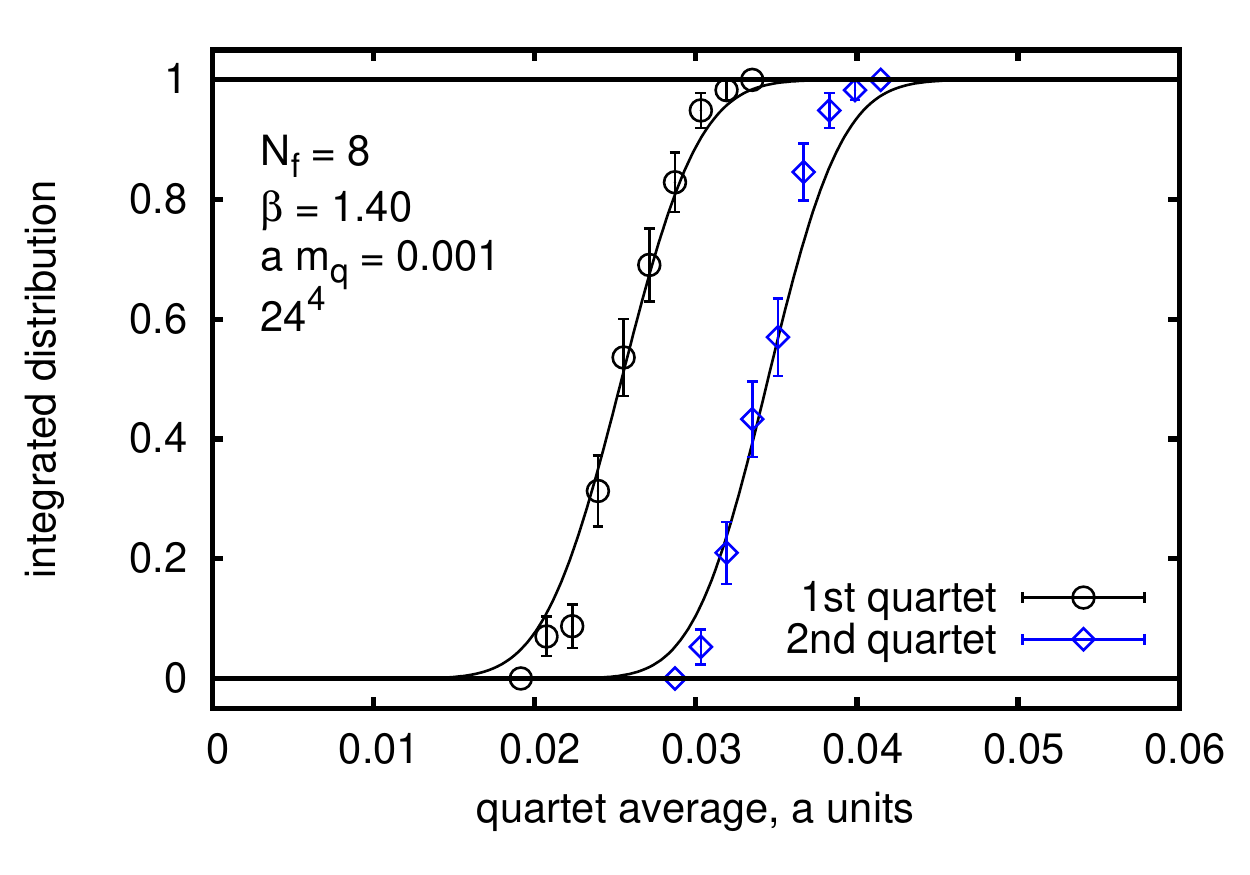}
\end{center}
\vskip -0.2in
\caption{The solid lines compare the integrated distribution of the two lowest
quartet averages to RMT predictions with $N_f=8$.}   
\label{fig:rmtNf8}
\vskip -0.1in
\end{figure}

The main qualitative features of the RMT spectrum are very similar in our
$N_f=8$ simulations as shown in Fig.~\ref{fig:rmtNf8}. One marked quantitative
difference is a noticeable slowdown in response to change in the coupling
constant.  As $\beta$ grows the recovery of the quartet degeneracy is
considerably delayed in comparison with the onset of p-regime Goldstone
dynamics. Overall, for the $N_f=4,8$ models we find consistency between the
p-regime analysis and the RMT tests.  Earlier, using Asqtad fermions at a
particular $\beta$ value, we found agreement with RMT even at $N_f=12$ which
indicated a chirally broken phase~\cite{Fodor:2008hn}. Strong taste breaking
with Asqtad fermions leaves the quartet averaging in question and the bulk
pronounced crossover of the Asqtad action as $\beta$ grows is also an issue.
Currently we are investigating the RMT picture for $N_f=9,10,11,12$ with our
much improved action with stout smearing. This action shows no
artifact transitions and handles taste breaking much more effectively. Firm
conclusions on the $N_f=12$ model to support our findings of $\chi{\rm SB}$
in the p-regime will require continued investigations.

\section{Inside the conformal window}

We start our investigation and simulations of the conformal window at $N_f=16$
which is the most accessible for analytic methods. We are particularly
interested in the qualitative behavior of the finite volume spectrum of the
model and the running coupling with its associated beta function which is
expected to have a weak coupling fixed point around $g^{*2} \approx 0.5$, as
estimated from the scheme-independent, two-loop beta
function~\cite{Heller:1997vh}.  A distinguished feature of the $N_f=16$
conformal model is how the renormalized coupling $g^2(L)$  runs with $L$, the
linear size of the spatial volume in a Hamiltonian or Transfer Matrix
description.  On very small scales the running coupling $g^2(L)$ grows with $L$
as in any other asymptotically free theory. However, $g^2(L)$ will not grow
large, and in the $L\rightarrow \infty$ limit it will converge to the fixed
point $g^{*2}$ which is rather weak, within the reach of perturbation theory.
There is non-trivial, small-volume dynamics which is illustrated first in the
pure gauge sector. 

At small $g^2$, without fermions,  the zero-momentum components of the gauge
field are known to dominate the
dynamics~\cite{'tHooft:1979uj,Luscher:1982ma,vanBaal:1986ag}.  With $SU(3)$
gauge group, there are twenty-seven degenerate vacuum states, separated by
energy barriers which are generated by the integrated effects of the non-zero
momentum components of the gauge field in the Born-Oppenheimer approximation.
The lowest-energy excitations of the gauge field Hamiltonian scale as $ \sim
g^{2/3}(L)/L$ evolving into glueball states and becoming independent of the
volume as the coupling constant grows with $L$.  Non-trivial dynamics evolves
through three stages as $L$ grows. In the first regime, in very small boxes,
tunneling is suppressed between vacua which remain isolated. In the second
regime, for larger $L$, tunneling sets in and electric flux states will not be
exponentially suppressed. Both regimes represent small worlds with
zero-momentum spectra separated from higher momentum modes of the theory with
energies on the scale of $2\pi/L$. At large enough $L$ the gauge dynamics
overcomes the energy barrier, and wave functions spread over the vacuum valley.
This third regime  is the crossover to confinement where the electric fluxes
collapse into thin string states wrapping around the box. 

It is likely that a conformal theory with a weak coupling fixed point at
$N_f=16$ will have only the first two regimes which are common with QCD. Now
the calculations have to include fermion
loops~\cite{vanBaal:1988va,Kripfganz:1988jv}.  The vacuum structure in small
enough volumes, for which the wave functional is sufficiently localized around
the vacuum configuration, remains calculable by adding in one-loop order the
quantum effects of the fermion field fluctuations.  The spatially constant
abelian gauge fields parametrizing the vacuum valley are given by $A_i(\vek
x)=T^aC^a_i/L$ where $T_a$ are the (N-1) generators for the Cartan subalgebra
of $SU(N)$. For $SU(3)$, $T_1=\lambda_3/2$ and $T_2=\lambda_8/2$.  With $N_f$
flavors of massless fermion fields the effective potential of the constant mode
is given by
\beq
V_{\eff}^{\vek k}(\vek C^b)=\sum_{i>j}V(\vek C^b[\mu^{(i)}_b-\mu^{(j)}_b])
-N_f\sum_{i}V(\vek C^b\mu^{(i)}_b+\pi\vek k),\label{eq:Vquark}
\eeq
with $\vek k=\vek 0$ for periodic, or $\vek k=(1,1,1)$, for  antiperiodic
boundary conditions on the fermion fields. The function $V(\vek C)$ is the
one-loop effective potential for $N_f=0$ and the weight vectors $\mu^{(i)}$ are
determined by the eigenvalues of the abelian generators.  For  SU(3)
$\mu^{(1)}= (1,1,-2)/\sqrt{12}$ and $\mu^{(2)}=\half(1,-1,0)$.  The correct
quantum vacuum is found at the minimum of this effective potential which is
dramatically changed by the fermion loop contributions. 
\begin{figure}[ht!]
\begin{center}
\begin{tabular}{cc}
\includegraphics[width=7.5cm,angle=0]{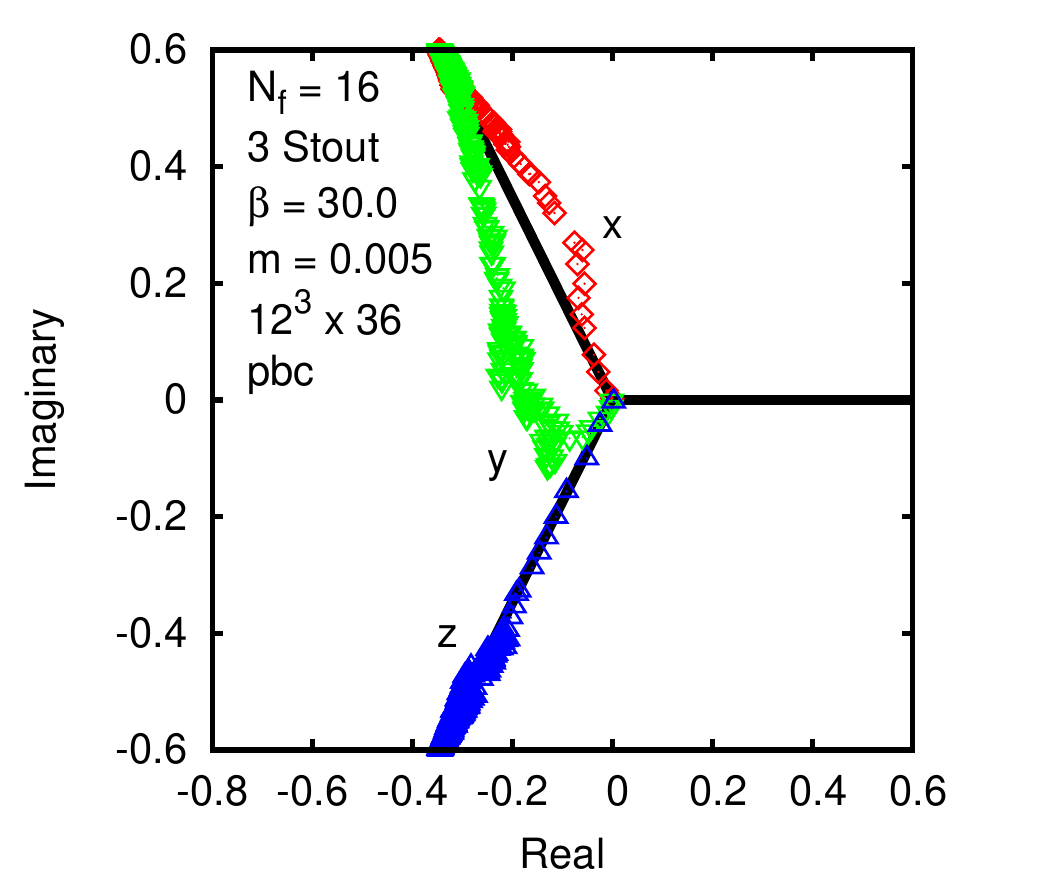}&
\includegraphics[width=7.62cm,angle=0]{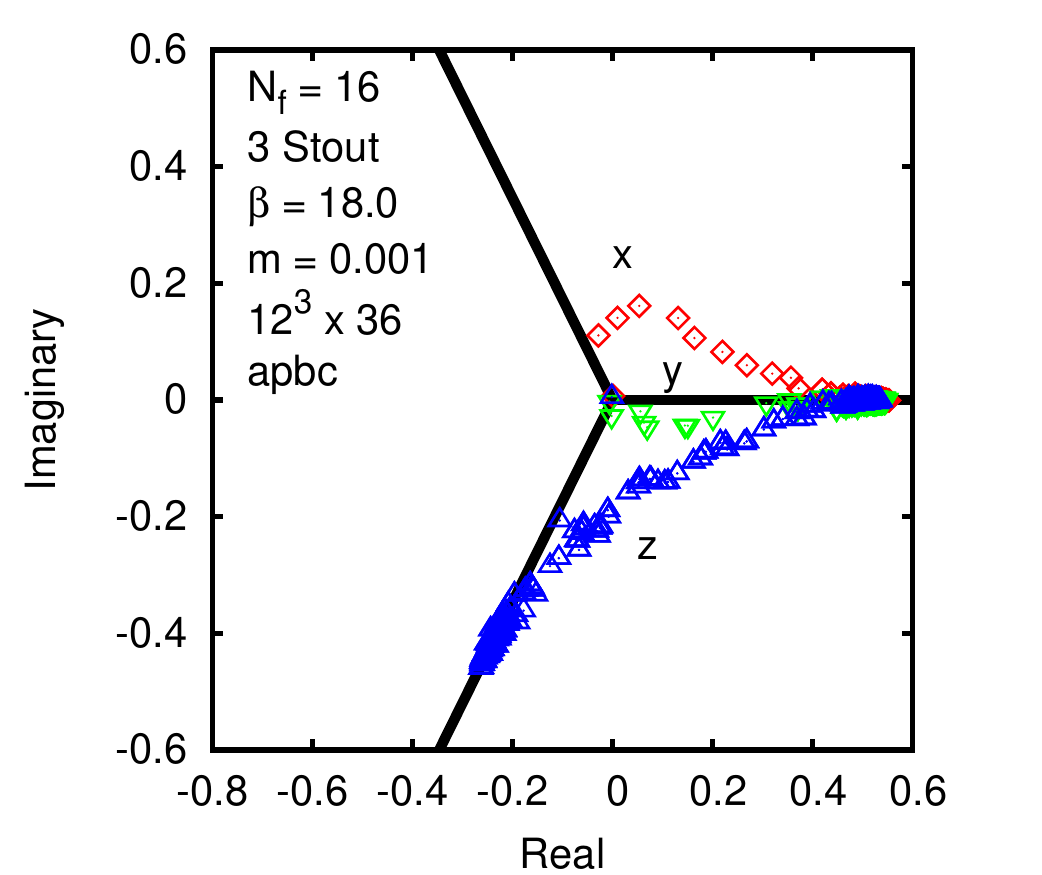}
\end{tabular}
\end{center}
\vskip -0.1in
\caption{The time evolution of complex  Polyakov loop distributions are shown 
from our $N_f=16$ simulations with $12^3\times 36$ lattice volume. Tree-level
Symanzik-improved gauge action is used in the simulations and staggered
fermions with three stout steps and very small fermion masses.}   
\label{fig:Polyakov}
\vskip -0.1in 
\end{figure}
The  Polyakov loop observables remain  center elements at the new vacuum
configurations with complex values; for $SU(N)$
\beq
P_j=\frac{1}{N}\tr\left(\exp(iC_j^bT_b)\right)=\frac{1}{N}\sum_n
\exp(i\mu^{(n)}_bC^b_j)=\exp(2\pi i l_j/N).\label{eq:PVac}
\eeq
This implies $\mu^{(n)}_b\vek C^b=2\pi\vek l/N$ (mod
$2\pi$), independent of $n$, and $V_{\eff}^{\vek k}=-N_f NV(2\pi\vek
l/N+\pi\vek k)$. In the case of antiperiodic boundary conditions, $\vek
k=(1,1,1)$, this is minimal only when $\vek l=\vek 0$ (mod $2\pi$). The quantum
vacuum in this case is the naive one, $A=0$ ($P_j=1$). In the case of periodic
boundary conditions, $\vek k=\vek 0$, the  vacua have $\vek l\neq\vek 0$, so
that $P_j$ correspond to non-trivial center elements.  For  SU(3),  there are
now 8 degenerate vacua characterized by eight different Polyakov loops,
$P_j=\exp(\pm 2\pi i/3)$.  Since they are related by coordinate reflections, in
a small volume parity (P) and charge conjugation (C) are spontaneously broken,
although CP is still a good symmetry~\cite{vanBaal:1988va}. 

Our simulations of the $N_f=16$ model below the conformal fixed point $g^{*2}$
confirm the theoretical vacuum structure. Fig.~\ref{fig:Polyakov} shows the
time evolution of Polyakov loop distributions monitored along the three
separate spatial directions. On the left side, with periodic spatial  boundary
conditions, the time evolution is shown starting from randomized gauge
configuration with the Polyakov loop at the origin. The system evolves into one
of the eight degenerate vacua selected by the positive imaginary part of the
complex Polyakov loop along the x and y direction and negative imaginary part
along the z direction. On the right, with antiperiodic spatial boundary
conditions, the vacuum is unique and trivial with real Polyakov loop in all
three directions. The time evolution is particularly interesting in the z
direction with a swing first from the randomized gauge configuration to a
complex metastable minimum first, and eventually tunneling back to the trivial
vacuum and staying there, as expected. The measured fermion-antifermion spectra
and the spectrum of the Dirac operator further confirm this vacuum structure.
Our plans include the continued  investigation of zero-mode gauge dynamics
which should clarify many important aspects of conformal and nearly conformal
gauge theories.

\vskip -0.1in
\section*{Acknowledgments}
We are thankful to Claude Bernard and Steve Sharpe for help with staggered
perturbation theory and to Ferenc Niedermayer for discussions on rotator dynamics.
We are
grateful to Sandor Katz and Kalman Szabo for the Wuppertal RHMC code, and for
some calculations, we used the publicly available MILC code.  We performed
simulations on the Wuppertal GPU cluster, Fermilab clusters under the auspices
of USQCD and SciDAC, and the Ranger cluster of the Teragrid organization.  This
work is supported by the NSF under grant 0704171, by the DOE under grants
DOE-FG03-97ER40546, DOE-FG-02-97ER25308, by the DFG under grant FO 502/1 and by
SFB-TR/55.

\end{document}